\documentclass[a4paper,fleqn,usenatbib]{mnras}

\usepackage[T1]{fontenc}
\usepackage{ae,aecompl}

\usepackage{graphicx}	
\usepackage{amsmath}	
\usepackage{amssymb}	
\usepackage{color}
\usepackage{mathptmx}
\usepackage{epsfig}
\usepackage{wasysym}
\usepackage{xfrac}
\usepackage{pifont}
\usepackage[table,xcdraw]{xcolor}
\usepackage{booktabs}
\usepackage{multirow}
\usepackage{comment}
\usepackage{subfigure}
\usepackage{longtable}

\title[Morphologies via unsupervised machine-learning]{Galaxy morphological classification in deep-wide surveys via unsupervised machine learning}

\author[G. Martin et al.]{
G. Martin,$^{1,2,3}$\thanks{E-mail: garrethmartin@arizona.edu} S. Kaviraj,$^{1}$ A. Hocking,$^{4}$  S. C. Read$^{1,5}$ and J. E. Geach$^{1}$\\
$^{1}$Centre for Astrophysics Research, School of Physics, Astronomy and Mathematics, University of Hertfordshire, College Lane, Hatfield AL10 9AB, UK,\\
$^{2}$Steward Observatory, University of Arizona, 933 N. Cherry Ave, Tucson, AZ, USA,\\
$^{3}$Korea Astronomy and Space Science Institute, 776 Daedeokdae-ro, Yuseong-gu, Daejeon 34055, Korea,\\
$^{4}$Centre for Computer Science and Informatics Research, University of Hertfordshire, Hatfield AL10 9AB, UK,\\
$^{5}$ I.N.A.F-Osservatotio Astronomico di Roma, via Frascati 33, 00040 - Monte Porzio Catone (Roma), Italy
}

\pubyear{2019}

\begin{document}
\label{firstpage}
\pagerange{\pageref{firstpage}--\pageref{lastpage}}
\maketitle


\begin{abstract}
Galaxy morphology is a fundamental quantity, that is essential not only for the full spectrum of galaxy-evolution studies, but also for a plethora of science in  observational cosmology (e.g. as a prior for photometric-redshift measurements and as contextual data for transient lightcurve classifications). While a rich literature exists on morphological-classification techniques, the unprecedented data volumes, coupled, in some cases, with the short cadences of forthcoming `Big-Data' surveys (e.g. from the LSST), present novel challenges for this field. Large data volumes make such datasets intractable for visual inspection (even via massively-distributed platforms like Galaxy Zoo), while short cadences make it difficult to employ techniques like supervised machine-learning, since it may be impractical to repeatedly produce training sets on short timescales. \textit{Unsupervised} machine learning, which does not require training sets, is ideally suited to the morphological analysis of new and forthcoming surveys. Here, we employ an algorithm that performs clustering of graph representations, in order to group image patches with similar visual properties and objects constructed from those patches, like galaxies. We implement the algorithm on the Hyper-Suprime-Cam Subaru-Strategic-Program Ultra-Deep survey, to autonomously reduce the galaxy population to a small number (160) of `morphological clusters', populated by galaxies with similar morphologies, which are then benchmarked using visual inspection. The morphological classifications (which we release publicly) exhibit a high level of purity, and reproduce known trends in key galaxy properties as a function of morphological type at $z<1$ (e.g. stellar-mass functions, rest-frame colours and the position of galaxies on the star-formation main sequence). Our study demonstrates the power of unsupervised machine learning in performing accurate morphological analysis, which will become indispensable in this new era of deep-wide surveys.
\end{abstract}




\section{Introduction}
The measurement of galaxy morphology is a fundamental topic in observational cosmology. Morphology is a strong function of the dynamical state of a galaxy, encodes the physical processes that dominate its evolutionary history \citep[e.g.][]{Martin2018b} and is strongly aligned with physical properties like stellar mass \citep[e.g.][]{Bundy2005}, star-formation rate \citep[e.g.][]{Bluck2014,Smethurst2015}, colour \citep[e.g.][]{Strateva2001,Skibba2009} and local environment \citep[e.g.][]{Dressler1980,Dressler1997,Postman2005}. For example, bulge-dominated galaxies typically have assembly histories that are richer in mergers \citep[e.g.][]{Conselice2006}, with the strength of the bulge correlating with the number of mergers \citep[e.g.][]{Hatton2003}. In comparison, the presence of a disc at the present day is a signature of a more quiescent formation history, with the buildup of stellar mass likely to be driven mainly by gas accretion and secular processes \citep{Codis2012,Kaviraj2014a,Martin2018c}. In a similar vein, at a given stellar mass, lower surface brightnesses or redder colours may indicate a larger role for tidal processes, like interactions and ram-pressure stripping, in the evolution of the galaxy in question \citep[e.g.][]{Dressler1980,Moore1999,Weisz2011,Martin2019}. Finally, morphological details, such as extended tidal features, are signposts of recent mergers and/or strong interactions \citep[e.g.][]{Kaviraj2014b,Kaviraj2019}, with the surface-brightness of these tidal features typically scaling with the mass ratios of the mergers in question \citep[e.g.][]{Peirani2010,Kaviraj2010}. 

In addition to its key role in the study of galaxy evolution, morphological information is useful for a vast array of astrophysical science. For example, it is used as a prior in photometric-redshift pipelines \citep[e.g.][]{Soo2018,Menou2018}, forms key contextual data in the classification of transient lightcurves \citep[e.g.][]{Djorgovski2012,Wollaeger2018} and is important for identifying the processes that trigger the onset of AGN activity in galaxies \citep[e.g.][]{Schawinski2014}. The measurement of accurate galaxy morphologies, particularly in large surveys which underpin our statistical endeavour is, therefore, a critical exercise.  

Over the past few decades a rich literature has emerged on methods for measuring galaxy morphology, especially in large observational surveys. These methods range from parametric techniques, which attempt to describe galaxy light profiles using small sets of parameters \citep[e.g.][]{Sersic1963,Simard2002,Odewahn2002,Lackner2012}, to non-parametric methods that reduce these light distributions to single values such as in the `CAS' system \citep[e.g.][]{Abraham1994,Conselice2003,Menanteau2006}, the Gini-M$_{20}$ coefficients \citep[e.g.][]{Lotz2004,Scarlata2007,Peth2016} or other non-parametric statistics such as the MID system \citep[e.g.][]{Freeman2013}. Recent work has increasingly harnessed the power of machine-learning to perform morphological analysis. Although the use of machine-learning in astronomy can be traced back at least as far as \citet{Lahav1995}, the recent literature has seen an explosion in the use of such techniques applied to a wide variety of problems in astrophysics \citep[e.g.][]{Huertas-Company2015,Ostrovski2017,Schawinski2017,Hocking2018,Goulding2018,DIsanto2018b,Siudek2018,An2018,Cheng2019,Ay2019}. 

While automated classification techniques, such as the ones described above, are particularly well-suited to efficiently processing large survey datasets, they are typically benchmarked against visual inspection \citep[e.g.][]{Kaviraj2010,Lintott2011,Simmons2017}, which produces arguably the most powerful and accurate measures of galaxy morphology. While time-consuming to perform, the development of the Galaxy Zoo \citep[GZ;][]{Lintott2011} platform has, in recent years, revolutionized the collection of visual classifications for large surveys. Using more than a million citizen-science volunteers, GZ has classified several contemporary surveys, like the SDSS and the HST legacy surveys \citep[e.g.][]{Lintott2011,Willett2017}. Automated methods, especially those that exploit machine-learning, have been routinely benchmarked against visual classifications from databases like GZ, and are now commonly deployed on large-scale survey data \citep[e.g.][]{Huertas-Company2015,Dieleman2015,Beck2018,Walmsley2019,Ma2019}. 

Notwithstanding the variety of techniques on offer, forthcoming `Big Data' surveys, e.g. from the LSST, present unprecedented challenges for performing morphological classification \citep{Robertson2017}. The sheer volume of data makes such surveys intractable for visual inspection, even via massively-distributed platforms like GZ. New techniques, which either combine visual and automated classification \citep[e.g.][]{Beck2018,Dickinson2019} or perhaps remove the need for visual classification altogether \citep[e.g.][]{Siudek2018,Hocking2018,Hendel2018,DIsanto2018a}, will be crucial in dealing with the unprecedented data volumes expected from these new surveys. The short cadence of rapidly-changing datasets, from instruments like the LSST, represents an additional hurdle and could make supervised machine-learning techniques challenging to deploy, as it may become impractical to repeatedly produce large, reliable training sets on short timescales, as the survey becomes progressively deeper. 

\textit{Unsupervised} machine-learning (UML) algorithms are ideally suited to the morphological analysis of Big Data surveys. Unsupervised techniques do not require visually-classified training sets and can, in principle, rapidly and autonomously compress an arbitrarily large galaxy population into a small number of `morphological clusters' comprised of galaxies with similar properties \citep[e.g.][]{Hocking2018}. These groups can then be benchmarked against visual classification which, if the number of groups is relatively small, becomes tractable even for individual researchers (and can be tackled easily using distributed systems like GZ). 

In this paper, we employ such a UML algorithm, which works by grouping pixels with similar properties and objects constructed from those pixels, like galaxies. Originally developed using Hubble Space Telescope (HST) data from the CANDELS \citep{Koekemoer2011} survey \citep{Hocking2017,Hocking2018}, we apply the algorithm to the Ultradeep layer of the Hyper Suprime-Cam Subaru-Strategic Program (HSC-SSP) Data Release 1 (DR1). We release a catalog of morphological classifications which can be used in conjunction with the HSC-SSP DR1 catalogue, explore the robustness of these classifications and discuss the applicability of the algorithm to surveys from forthcoming instruments like LSST (whose deep-wide-fast dataset will reach the same depth as the HSC-SSP Ultradeep survey after $\sim$10 years). We also plan to release catalogues of morphological classifications based on both forthcoming HSC-SSP data releases and datasets from future instruments like LSST will be released as data becomes available. 

This paper is structured as follows. In Section \ref{sec:HSC}, we describe the unsupervised graph-clustering algorithm that underpins this study. In Section \ref{sec:algorithm}, we outline the properties of the HSC-SSP and the ancillary data used in this study. In Section \ref{sec:catalogue}, we describe the benchmarking of the algorithm outputs using visual classification, the completeness of the resultant morphological catalogue and the contents of the released data products. In Section \ref{sec:analysis}, we explore the robustness of the classifications, by comparing the properties of galaxies in different morphological groups to known trends in these properties as a function of morphology, at $z<1$. We summarise our results in Section \ref{sec:summary}. 


\section{The Hyper Suprime Cam Subaru Strategic Program (HSC-SSP)}
\label{sec:HSC}

\subsection{Survey description}
The HSC-SSP \citep{Aihara2018a} is a multi-layered imaging survey in $grizy$ (and 4 narrow-band filters), using the Hyper Suprime-Cam \citep[HSC,][]{Miyazaki2012} on the 8.2m Subaru Telescope. HSC has a 1.5 degree diameter field of view and a 0.168 arcsec pixel scale, with a median $i$-band seeing of $\sim$0.6 arcsec. The survey, which began in 2014, is being carried out using 300 nights over 5-6 years. The fields are chosen to be low in Galactic dust extinction and to have overlap with several well-known multi-wavelength data-sets, including SDSS/BOSS \citep{Eisenstein2011}, X-ray surveys from XMM \citep{Jansen2001} and eROSITA \citep{Merloni2012} and near-/mid-infrared imaging surveys e.g. VIKING/VIDEO \citep{Jarvis2013} and UKIDSS \citep{Lawrence2007}.

The final HSC-SSP dataset (expected in 2021) will provide three layers: a `Wide' layer covering an area of 1400 deg$^2$ with a target $i$-band depth of 26.2 mag arcsec$^{-2}$, a `Deep' layer covering an area of 27 deg$^2$ with a target $i$-band depth of 27.1 mag arcsec$^{-2}$ and an `Ultradeep' layer covering an area of 3.5 deg$^2$ with a target $i$-band depth of 27.7 mag arcsec$^{-2}$ \citep{Aihara2018a}. The layers are nested, so that the Ultradeep layers are included in the Deep fields and the Deep regions are included in the Wide fields. 

Here, we use the HSC-SSP DR1\footnote{https://hsc.mtk.nao.ac.jp/ssp/data-release/}, which has released 108~deg$^2$, 26~deg$^2$, and 3.5~deg$^2$ in the Wide, Deep and Ultradeep layers, with current depths of $i \sim$ 26.4, $\sim$26.5, and $\sim$27. mag, respectively ($5\sigma$ for point sources) \citep{Aihara2018b}. The survey is split into a number of 1.5~deg wide square `tracts', each covering approximately a single HSC pointing. Each tract is further separated into $9\times9$ patches, consisting of approximately $4200\times4200$ pixels. Here we use stacked, sky-subtracted images, with WCS co-ordinate corrections applied and calibrated magnitude zero-points. 


\subsection{Data}

For object centroids and observed photometry, we use \texttt{cModel} magnitudes \citep[][]{Stoughton2002}, which are released as part of the HSC-SSP DR1 \texttt{forced} catalogue. These are computed using the HSC-SSP reduction pipeline, using the $i$-band as the primary reference wavelength. We additionally calculate surface brightnesses using the Kron radius, by dividing the flux within this radius by the area of the aperture. 


In order to infer physical properties and photometric redshifts for the galaxies in our sample, we use results from the \textsc{mizuki} \citep{Tanaka2015} template-fitting code, that have been released as part of the HSC-SSP~DR1 \citep{Tanaka2018}. Redshifts are derived purely from HSC $g$, $r$, $i$, $z$ and $y$ band \texttt{cModel} magnitudes, for all primary objects detected in at least three bands. The \textsc{mizuki} code uses spectral energy distribution (SED) fitting to templates generated from the \citet{Bruzual2003} stellar population synthesis models, in order to self-consistently estimate redshifts and physical properties of individual galaxies. Redshift-dependent Bayesian priors are applied to physical parameters like stellar mass and the star-formation rate (SFR). 
We use values from the public HSC-SSP DR1 \texttt{photoz\_mizuki} catalogue for photometric redshifts, SFRs, stellar masses and rest-frame magnitudes and colours. 
For full details of the HSC-SSP DR1 we direct readers to \citet{Aihara2018b}.


\section{An unsupervised algorithm for morphological classification based on the clustering of graph-based representations }
\label{sec:algorithm}

The graph-clustering algorithm that underpins this study is described in detail in \citet{Hocking2017} and \citet{Hocking2018}. The primary application of this algorithm is to provide a means of efficiently classifying large quantities of unseen data into small groups of visually similar objects, so that the groups can be benchmarked against visual classifications. The ability to do so quickly, without relying on pre-existing training sets, is essential in the LSST era, as the high survey cadence, rapidly evolving data and the new parameter spaces it will explore will make it challenging to construct the unbiased training sets that will be required.

The technique has been previously tested on other astronomical datasets. \citet{Hocking2018} has demonstrated this method applied to HST data from CANDELS. In this work, the algorithm has been validated by showing that the characteristics of objects, separated into different morphological clusters, show consistency in terms of their basic properties and structural parameters (magnitude, M$_{20}$, colours), as well as strong concordance with classification data from GZ (e.g. smoothness, disciness etc). Separately, the algorithm has been shown to efficiently separate ellipticals and spirals, when benchmarked against expert human classifiers, and shows promise in identifying lensed galaxies \citep{Hocking2017}.

In the following sections, we describe the main components of the algorithm. 

\subsection{Feature selection}
\label{sec:feature_selection}

\begin{table}
\caption{Parameters used for the feature extraction step.}
\begin{tabular}{lll}
  \toprule Parameter & Description & Value\\ 
  \midrule
  $r$ & Side length of a square sub-image patch in pixels & 16 \\
  $n$ & Number of bins in the radial power spectrum & 8 \\
\bottomrule
\end{tabular}
\label{table:FE}
\end{table}

The ultimate aim of the method is to automatically identify different groups of galaxies using HSC pixel data. Although the source data may be used directly, it is more useful to transform the data in a way that removes any irrelevant information. As our aim is to morphologically classify galaxies, we transform the data so that irrelevant information like galaxy orientation is removed.

Importantly, the selected features must avoid redundancies or over-fitting, and also remain invariant to galaxy rotation, scale and orientation. Simplicity in the features selected is also desirable, in order that the results of the algorithm remain human-interpretable. Below, we outline the feature extraction process, which is also described in \citet{Hocking2017} and \citet{Hocking2018}.

We first extract $r\times r$ pixel sub-image patches around each detected pixel in each HSC tract, where $r$ is the patch size (Table \ref{table:FE}). In order to reduce the time that the algorithm takes to run, and to avoid including pixels that contain no useful information, we only extract pixels that are $1\sigma$ above the noise level, determined by a simple sigma-clipping \citep[e.g. as implemented by][but see Appendix \ref{app:cross_match} for a discussion of potential improvements]{Bertin1996, Robitaille2013}.  

Following the initial detection and extraction step, we produce a rotationally invariant representation of each patch, by evaluating the radially averaged pixel intensity power spectrum, with $n$ bins (Table \ref{table:FE}) for each of the five bands ($g$, $r$, $i$, $z$, $y$). This is done by first calculating the 2D Fast Fourier transform \citep[FFT;][]{Ballard1982} for each patch and then multiplying it by its conjugate. The zero frequency component is then rearranged to the centre of the 2D matrix and the azimuthally averaged radial profile calculated. It is important that the patch size is large enough to sample the spatial scales over which the data varies (e.g. that it is larger than the PSF). Other common feature representations, including Rotationally Invariant Feature Transform (RIFT), spin intensity images \citep{Lazebnik2005} and histogram of gradient \citep[HoG;][]{Birk1979} were trialed \citep{Hocking2017}, with the power spectrum representation found to have the best performance and efficiency for separation of late-type and early-type galaxies, when compared to human expert classifications of CANDELS data.

Each $n$ element power spectrum is concatenated into a $5\times n$ element feature vector, $\mathbf{p}$, which effectively encodes pixel intensity, colour and spatial frequency information for each sub-image patch in a rotationally invariant manner. Each feature vector is then combined into a patch data matrix, $\mathbf{P}$, which contains the feature vectors for every patch. Table \ref{table:FE} presents the values of $r$ and $n$ used for this feature extraction step. 

We note briefly that an alternative approach to the \textit{engineered} feature description detailed above, where the optimum feature representation is instead \textit{learned} \citep[e.g.][]{Coates2011,Cheriyadat2013,Tao2015} could also be used. While it is possible that this approach could produce better results, such an approach would also produce outputs that are difficult to interpret. Additionally, this could have significant implications for the speed of the algorithm, as it would introduce an additional learning step and also increase the dimensionality of the feature space that must be explored in later steps. This could potentially significantly increase the time required to produce classifications. Since we aim to be able to quickly and repeatedly produce classifications for rapidly changing datasets (e.g. from the LSST), and produce outputs where each step can be easily validated by human inspection, slowing down the algorithm and obfuscating the outputs in not desirable, even if it produces an improvement in the quality of the classifications.


\subsection{Feature extraction}
\label{sec:clustering}

The next step is to use clustering methods to learn an accurate topological map (model) of the patch data matrix, $\mathbf{P}$, and then sub-divide the nodes within this map into coarser groups of feature vectors, thus producing a library of distinct `patch types'.


\subsubsection{Growing neural gas}

We use a growing neural gas \citep[GNG,][]{Fritzke1995} algorithm to learn the optimal representation of the data, based on the patch data matrix, $\mathbf{P}$. The data are first normalised over each feature component to ensure homoscedasticity, preventing undue influence from components with relatively high variance.  The GNG algorithm then produces a graph representation of the data, by iteratively growing a graph of nodes with \textit{topological} neighbouring nodes in the graph connected by edges. The result is a topology-preserving map with an induced Delaunay triangulation \citep{Okabe2009}. Edges that are no longer part of the induced Delaunay triangulation must, however, be removed. This is achieved by removing edges that have reached a given \textit{age}, $a_{max}$, without being connected to another node. The GNG algorithm is applied to $\mathbf{P}$ using the following steps:

\begin{table}
\caption{Parameters used for the growing neural gas (GNG)s, hierarchical clustering (HC)s and morphological classification steps.}
\begin{tabular}{lll}
  \toprule Parameter & Description & Value\\ 
  \midrule
  $N$ & Maximum number of nodes in the graph & 200,000 \\
  $\lambda$ & Samples processed before new node added & 100 \\
  $a_{max}$ & Maximum age before an edge is removed & 50 \\
  $\epsilon_{b}$ & Size of the adjustment in step (i) & 0.2 \\
  $\epsilon_{n}$ & Size of adjustment for neighbours in step (i) & 0.006 \\
  $\alpha$ & Error reduction to node with the largest error & 0.5 \\
  $\beta$ & Error reduction to all nodes & 0.995\\
  $N_{g}$ & Target number of HC groups & 1500\\
  $k$ & Number of groups produced by $k$-means & 160\\
\bottomrule
\end{tabular}

\label{table:GNG}
\end{table}

\begin{enumerate}
\item First, two nodes are initialized with positions using two randomly selected feature vectors from the patch data matrix, $\mathbf{P}$. Each node is, therefore, located within a $5\times n$ dimensional feature space with the same dimensionality as the number of elements of $\mathbf{p}$. A new random feature vector, $\mathbf{p}'$, is then drawn from $\mathbf{P}$ and the following steps applied:

\begin{itemize}

\item  The two nearest nodes to the feature vector, whose positions in the feature space we designate $\mathbf{s}_{0}$ and $\mathbf{s}_{1}$, are identified such that the Euclidean distance from $\mathbf{p}'$, is minimised. $\mathbf{s}_{0}$ is the closest node to $\mathbf{p}'$ and $\mathbf{s}_{1}$ is the second closest.

\item If an edge connecting the two nodes, $\mathbf{s}_{0}$ and $\mathbf{s}_{1}$, does not already exist it is created. The two connected nodes are called \textit{topological} neighbours. Whenever two nodes are connected by an edge, the edge is also assigned an age, $a$, which is initially set to $0$, and the age of all other edges connected to $s_0$ are incremented by $1$.

\item The closest node to $\mathbf{p}'$, $\mathbf{s}_{0}$, is assigned an error equal to the square of their separation:

\begin{equation}
\sigma(\mathbf{s}_{0}) = ||\mathbf{p}' - \mathbf{s}_{0}||^{2}.
\end{equation}

\item $\mathbf{s}_{0}$ and its direct topological neighbours (i.e. those directly connected by edges) are all moved towards $\mathbf{p}'$ by a fraction ($\epsilon_{b}$ and $\epsilon_{n}$ respectively) of their separation from $\mathbf{p}'$, thus causing adaptation of the map towards the input data:

\begin{equation}
    \begin{array}{l}
    \Delta \mathbf{s}_{0} = \epsilon_{b} (\mathbf{p}' - \mathbf{s}_{0})\\  
    \Delta \mathbf{s}_{n} = \epsilon_{n} (\mathbf{p}' - \mathbf{s}_{n}).
  \end{array}
\end{equation}


\item All edges with ages larger than the maximum age (where $a > a_{max}$) are removed. Any nodes that no longer have topological neighbours are also removed.
\end{itemize}

\item This procedure is repeated until $\lambda$ feature vectors have been processed, after which:

\begin{itemize}
\item A new node, $\mathbf{s}_{r}$, is inserted at the mid-point between the node with the highest error, $\mathbf{s}_{q}$, and its highest error topological neighbour, $\mathbf{s}_{f}$.

\item The edges connecting the two nodes are removed and new edges are created connecting $\mathbf{s}_{q}$ and $\mathbf{s}_{f}$ to $\mathbf{s}_{r}$.

\item The error of $\mathbf{s}_{q}$ and $\mathbf{s}_{f}$ is decreased by multiplying their errors with the parameter, $\alpha$, and the error of $\mathbf{s}_{r}$ is initialised with the same error as $\mathbf{s}_{q}$.

\item The error of every node is decreased by multiplying their errors with the parameter $\beta$.
\end{itemize}

\item This is continued until the stopping criterion is met (i.e. $N$ nodes has been reached).

\end{enumerate}

The accumulation of errors in step (ii) ensures that the algorithm places new nodes in areas of the parameter space where the mapping from the model to the data is poor. Once the stopping criterion is met, we take a matrix containing the final positions of all the nodes within the feature space, $\mathbf{N}$, as the output. Table \ref{table:GNG} presents the values of the parameters used for the GNG step. We note that the exact value of these parameters is not important for the outcome, but does affect the time it takes for the graph to converge. Any sensible choice of parameters will always result in adaptation towards the input data, but a poor choice of parameters may result in inefficient performance, requiring a large number of iterations to finish.


\subsubsection{Hierarchical clustering}

Agglomerative (`bottom-up') hierarchical clustering \citep[HC;][]{Johnson1967} of the GNG output is used to produce a hierarchical representation of the nodes in the topological map. At each iteration, the HC algorithm initially tries to cluster the most similar nodes into pairs, with similarity measured, in this case, by the Pearson correlation. The Pearson correlation between the nodes $\mathbf{a}$ and $\mathbf{b}$ is calculated by comparing their position vectors and given by their co-variance divided by the product of their standard deviations:

\begin{equation}
\label{eqn:Pearson}
\rho(\mathbf{a},\mathbf{b}) = \frac{\mathrm{cov}(\mathbf{a},\mathbf{b})}{\sigma_{\mathbf{a}}\sigma_{\mathbf{b}}},
\end{equation}

where $\mathrm{cov}(\mathbf{a},\mathbf{b})$ is the co-variance between the two node position vectors, given by $\Sigma_{i}^{n}(\mathbf{a}_{i}-\overline{\mathbf{a}})(\mathbf{b}_{i}-\overline{\mathbf{b}})$ and $\sigma_{\mathbf{a}}$ and $\sigma_{\mathbf{b}}$ are simply the standard deviation of each position vector, given by $\sqrt{\Sigma_{i}^{n}(\mathbf{a}_{i}-\overline{\mathbf{a}})^{2}}$ and $\sqrt{\Sigma_{i}^{n}(\mathbf{b}_{i}-\overline{\mathbf{b}})^{2}}$ respectively.

At each subsequent iteration the algorithm merges clusters into pairs of similar clusters and so on, until only a single cluster remains. A particular advantage of this method is that it enables us to select the desired level of detail that we use to segment the GNG graph. The clusters can have disparate sizes and separations and therefore the method makes no assumptions about the structure of the data.


\subsection{Constructing feature vectors}
\label{sec:feature_vectors}

After a library of patch types has been produced from a subset of the data by the GNG algorithm and then reduced via HC, it is possible to construct \textit{object} feature vectors. Individual patches must be assembled into objects, either using existing detection maps or, as we use in this case, connected component labeling algorithms \citep[e.g.][]{Galler1964}. Each of the patches is assigned a patch type, based on the closest node defined in the previous step. They can then be described using a histogram of patch types i.e. an object feature vector. The feature vector describes the frequency of different patch types that the object consists of, thereby encoding an easily manipulated description of that object. 

\begin{figure}
    \centering
    \includegraphics[width=0.45\textwidth]{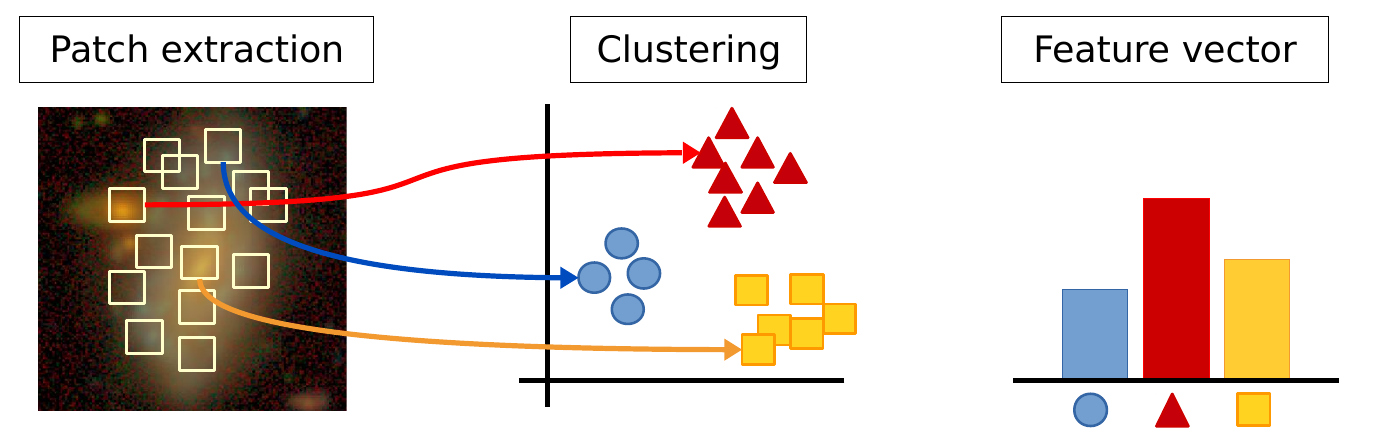}
    \caption{A schematic view of the morphological classification process. Patches are extracted around detected pixels in survey images and clustering methods are used to group these patches into a library of patch types. Galaxy feature vectors can then be constructed by creating a histogram for each object which describes the frequency of each patch `type'.}
    \label{fig:cartoon}
\end{figure}

The number of groups that patch types are clustered into, and therefore the length of the feature vector, can be changed according to the complexity of the data that is being classified. In this case it has a value $N_{g}=1500$, equal to the number of clusters produced by the HC algorithm. The feature vector of an object should, therefore, encode the basic visual characteristics of that object, making it possible to identify visually similar objects. Fig \ref{fig:cartoon} illustrates the process of extracting patches from multi-band survey data (Section \ref{sec:feature_selection}), assembling a library of patch types (Section \ref{sec:clustering}) and, finally, constructing feature vectors for each object (Section \ref{sec:feature_vectors}).

\subsection{Producing morphological clusters}

In order to finally classify galaxies into morphological clusters, we first define the similarity between feature vectors. Again, we use the Pearson correlation (Equation \ref{eqn:Pearson}), in order to define the distance between feature vectors in this new feature space, although other distance measures (e.g. Euclidean distance or cosine distance) may be used and may accentuate different features. Additionally, we apply `term frequency-inverse document frequency' \citep[TF$*$IDF,][]{Rajaraman2011} weightings when calculating the distance, in order to increase the importance of patch types with the greatest discriminatory power, and reduce the importance of patch types that are relatively common between all objects.

Once we have produced a feature vector that encodes the visual characteristics of each object, and defined a distance measure for these feature vectors, it is finally possible to group these objects by their visual similarity. This can be done either by direct comparison, or a similarity search, of individual feature vectors e.g. searching for other objects that are most similar, or closest in the feature space, to the feature vector of a given object, or by applying a clustering algorithm to the feature vectors in order to group them. 

In order to ensure cleaner classifications, we exclude any objects that are comprised of fewer than $15$ pixels. Using $k$-means clustering \citep[e.g.][]{MacQueen1967}, we separate our object feature vectors into $k$ morphological clusters (in this paper we have chosen $k$ to be 160, which is intended to demonstrate the ability of the algorithm to distinguish subtle visual differences, however arbitrary values of $k$ may be chosen). We calculate silhouette scores for the objects in each morphological cluster, in order to evaluate the overall quality of the clustering, as well as the correspondence of individual objects to the average properties of the group they are assigned to. The silhouette score for a given object is calculated based on the level of similarity to both its corresponding cluster and all other clusters \citep[see][]{Rousseeuw1987}. Silhouette scores range from -1 to 1: a high silhouette score indicates that the object is well matched to its own cluster and distinct from neighbouring clusters. 
For more efficient visual classification it may be desirable to select values of $k$ that so that the mean silhouette score is optimised (see Section \ref{sec:release}).
We calculate silhouette scores for individual objects, as well as averages on a group and global level, which are included in the catalogues presented in Section \ref{sec:release}. With $k=160$, we obtain a global mean silhouette score of 0.26.

Using the parameters described above, the algorithm takes around 40~ms per pixel in order to perform feature extraction, generate a model from training data and perform the classification. Feature extraction and classification, using an existing model applied to unseen data, takes only around 1-2~ms per pixel, on a single thread of execution on a contemporary desktop computer with an Intel CPU. The feature extraction and classification steps can be easily split up and executed concurrently \citep{Herlihy2011}. This property makes the algorithm efficient, even on very large volumes of data (e.g. surveys from instruments like SDSS or LSST). 

The algorithm is implemented in a combination of C\#, utilising .NET Core 2 libraries and Python, relying on the numpy, astropy and scikit-learn modules (which are implemented in a mixture of Python, C and Cython). Even without parallelization of the extraction and classification steps, the algorithm performs well on large datasets. For example, the entire 3.5~deg$^{2}$ of the HSC Ultradeep dataset used in this paper was processed in under 40~CPU~hours, including feature selection, extraction and classification. Scaling up to much larger data volumes will also be possible. For example, under the conservative assumption that 1 per cent of the approximately $10^{12}$ pixels that make up the SDSS are detected, the entirety of SDSS could be processed in under 3000~CPU~hours (assuming a modeling/feature extraction step has already been performed). Assuming the same set of assumptions for LSST (although LSST images will have more detected pixels than SDSS due to greater depth), the smaller pixel size and larger area of LSST would require around 16,000 CPU hours. This still represents a relatively trivial amount of processing time on even a modest high performance computing (HPC) cluster (e.g. around 1 day with 500 threads of execution).


\subsection{Cross-matching objects from the algorithm with the HSC-SSP}

We cross-match the galaxy centroids from the HSC-SSP DR1 Ultradeep catalogue with the object centroids from the graph-clustering algorithm, excluding objects that do not have a match within $0.8''$, which is approximately the PSF of the worst HSC $i$-band seeing (note that the median $i$-band seeing is $0.6''$). Of the 89,257 objects for which the algorithm produced classification, 53,003 have more than 15 pixels, which we consider to be large enough for reliable classification, as a sufficient range of spatial scales can be captured. Of these, 41,062 (77 per cent) have centroids that match an object in the HSC catalogue within $0.8''$. Mismatch between centroids arises because, at present, we use a simple connected component labelling algorithm to identify objects, rather than the individual segmentation maps used by the HSC-SSP pipeline. The mismatch becomes increasingly worse for very large objects (see Fig \ref{fig:npix}) and is, therefore, principally a problem in the very local Universe. However, in our analysis below, we study more distant objects ($z \gtrsim 0.3$), with much smaller projected sizes. The fraction of matched objects is, therefore, much larger as, on average 95 per cent of objects smaller than 100 pixels are successfully cross-matched (compared with 15 per cent of objects larger than 1000 pixels).


\section{Morphological catalogue}
\label{sec:catalogue}

\subsection{Benchmarking of morphological clusters via visual classification}

\label{sec:classification}

\begin{figure}
    \centering
    \subfigure[\textit{Spiral galaxies.}]{\includegraphics[width=0.45\textwidth]{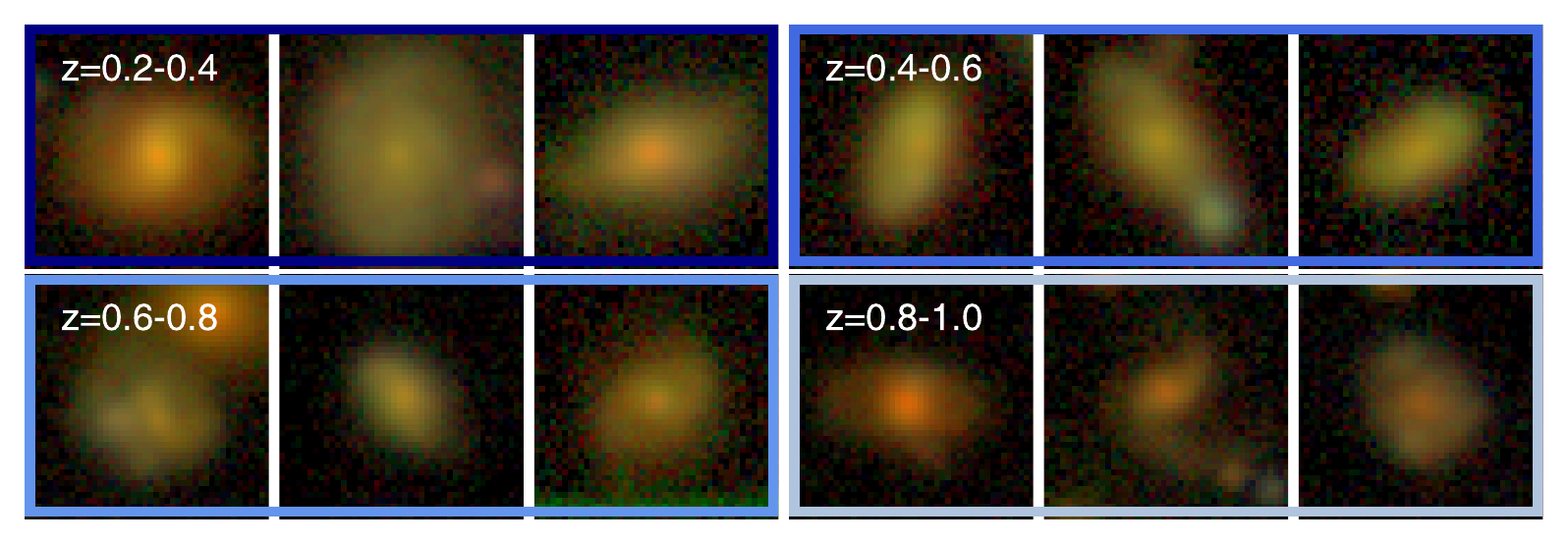}}
    \subfigure[\textit{S0/Sa galaxies.}]{\includegraphics[width=0.45\textwidth]{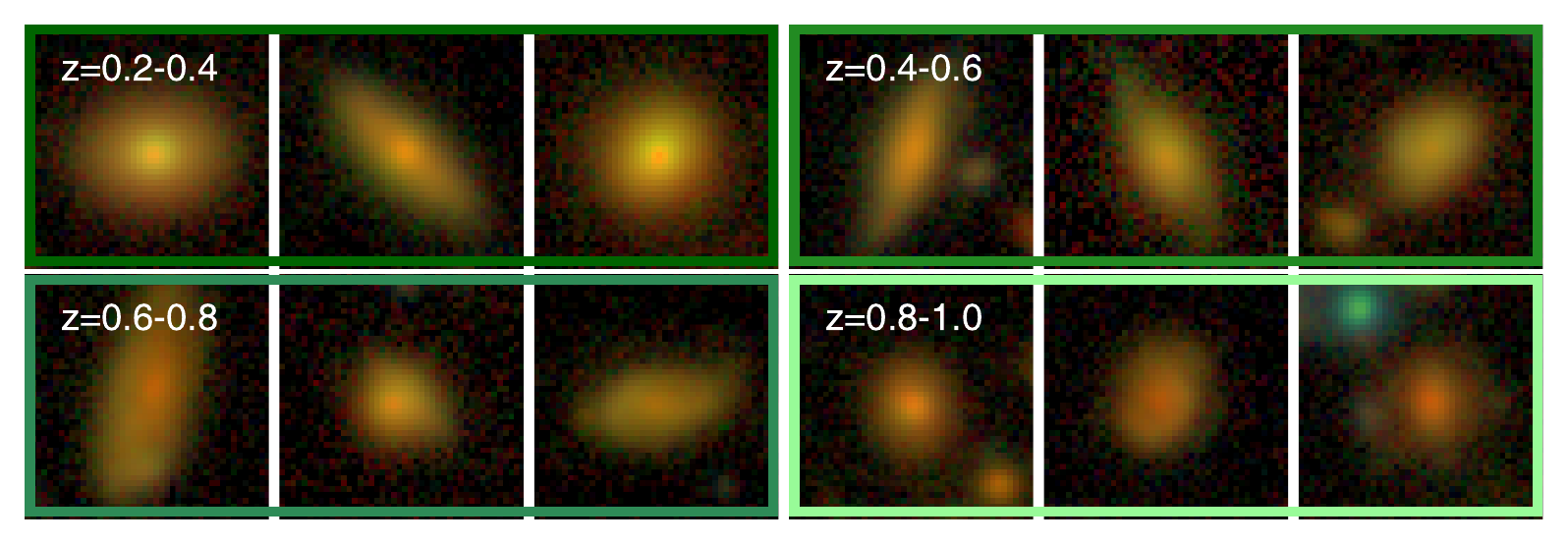}}
    \subfigure[\textit{Elliptical galaxies.}]{\includegraphics[width=0.45\textwidth]{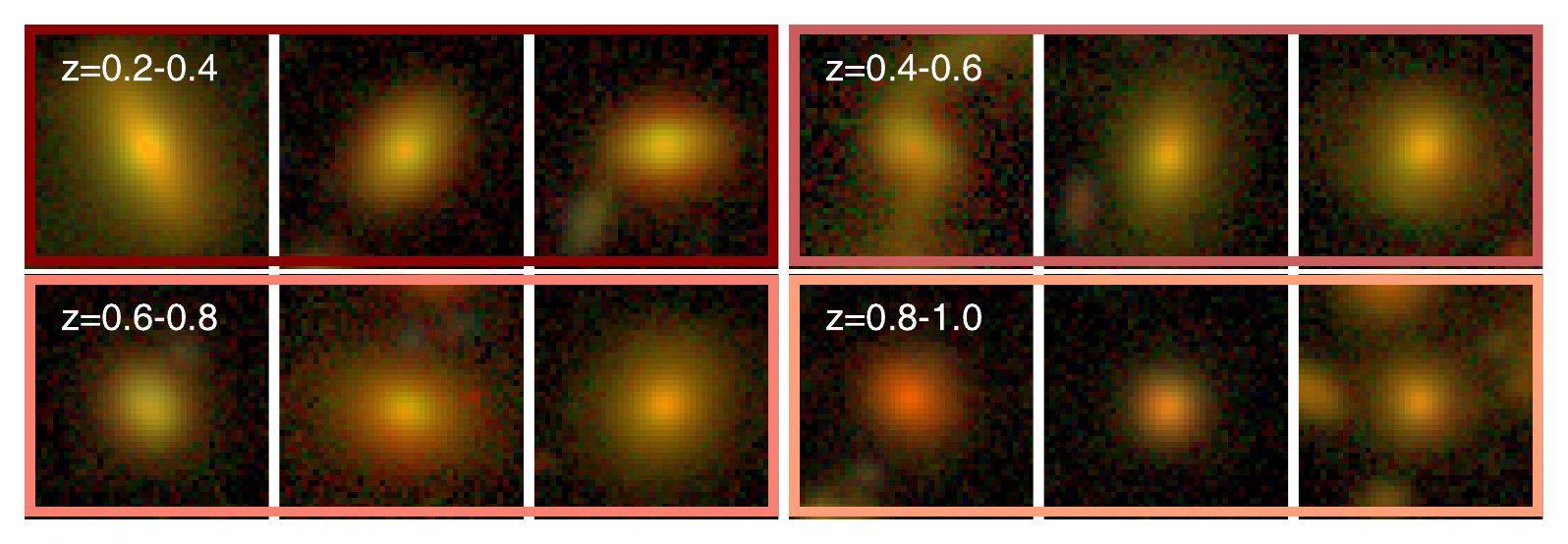}}
    \caption{$g$-$r$-$i$ false colour images showing a random selection of galaxies from each major morphological group. The samples are further split into bins of redshift, indicated by the label in the top right of each coloured box. Panel (a) shows objects classified as spirals, panel (b) shows objects classified as S0/Sa and panel (c) shows objects classified as ellipticals.}
    \label{fig:samples}
\end{figure}

The unsupervised graph-clustering algorithm allows us to effectively compress the galaxy population into a small number of morphological clusters. Crucially, the number is small enough to make visual classification of these clusters tractable for individual researchers. To generate a usable morphological catalogue, we benchmark the outputs of the algorithm via visual classification of each of the $k=160$ morphological clusters. These classifications are based on a subset of $g$-$r$-$i$ images of the 10 highest silhouette-score objects in each cluster, plus a sample of 10 objects selected at random, in order to assess the morphological purity of the cluster. We do not classify individual galaxies but perform visual classification on the cluster as a whole. 

We classify each cluster into one of three broad Hubble \citep{Hubble1936} morphological types: elliptical galaxies, S0/Sa galaxies and spiral galaxies. We also store finer morphological information, e.g. the type of spiral morphology (Sb, Sc, Sd) and noteworthy colour or structural features (e.g. when spirals appear unusually red or show clumpy structure, or when elliptical galaxies appear unusually blue). Except in Section \ref{sec:star_separation}, we only consider objects which are extended \citep[based on the difference between the PSF magnitude and the \texttt{cModelMag} magnitude;][]{Eisenstein2001}. We indicate the total number of objects in each morphological cluster that are not extended in Table \ref{table:props}. Figure \ref{fig:samples} shows a random selection of objects that are classified as having spiral, S0/Sa and elliptical morphologies, split into four redshift ranges. Note that, although a sample of individual objects in each cluster are visually classified in order to determine a morphological type for that cluster, the majority of objects in each cluster are unseen. 

\begin{figure}
    \centering
    \subfigure[Cluster 10: \textit{Sc/Sd galaxies.}]{\includegraphics[width=0.43\textwidth]{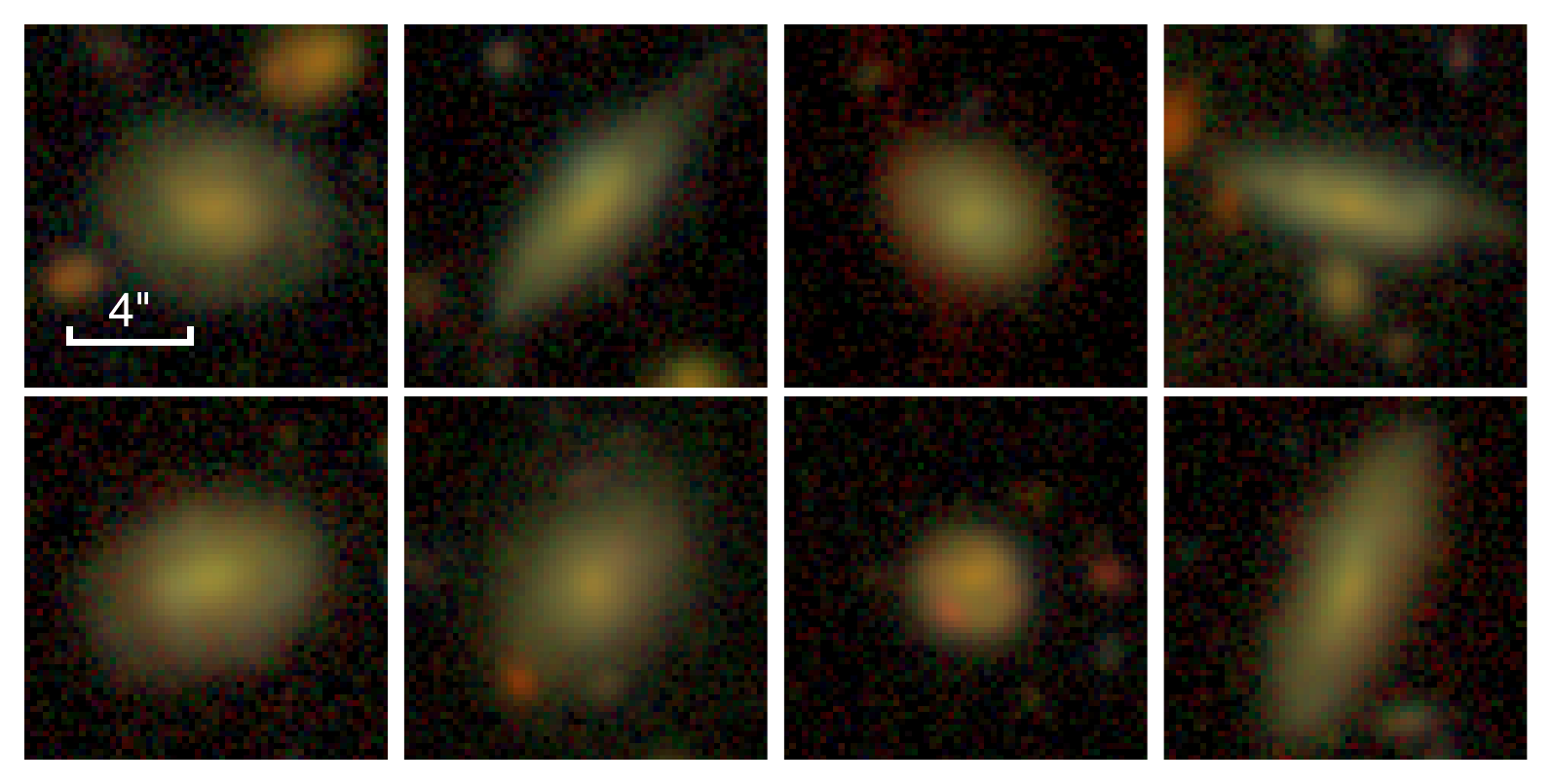}\label{fig:groups:a}}
    \subfigure[Cluster 14: \textit{High-$z$ mergers.}]{\includegraphics[width=0.43\textwidth]{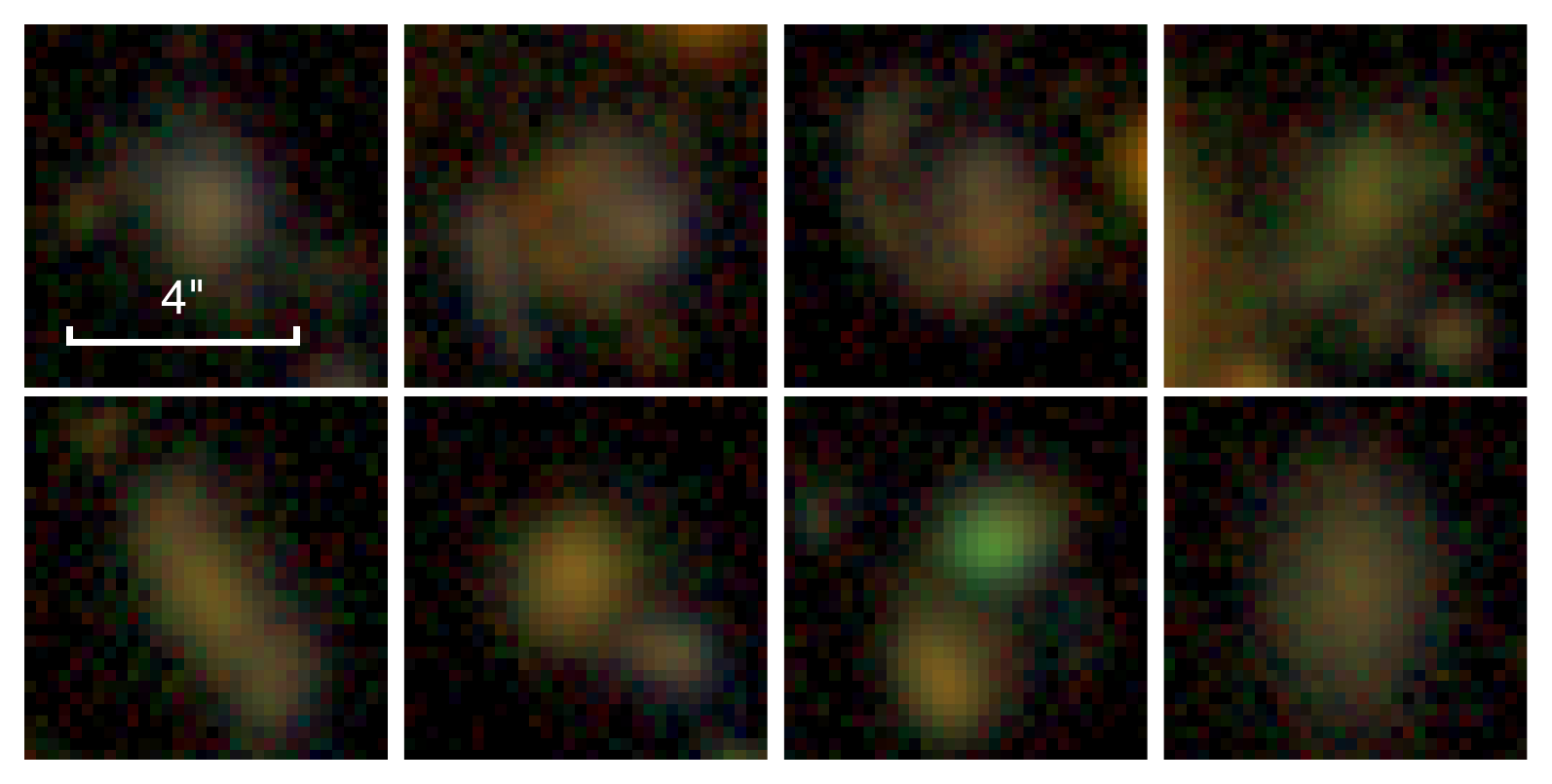}\label{fig:groups:b}}
    \subfigure[Cluster 122: \textit{Discs with blue rings, possibly indicative of the recent accretion of blue satellites.}]{\includegraphics[width=0.43\textwidth]{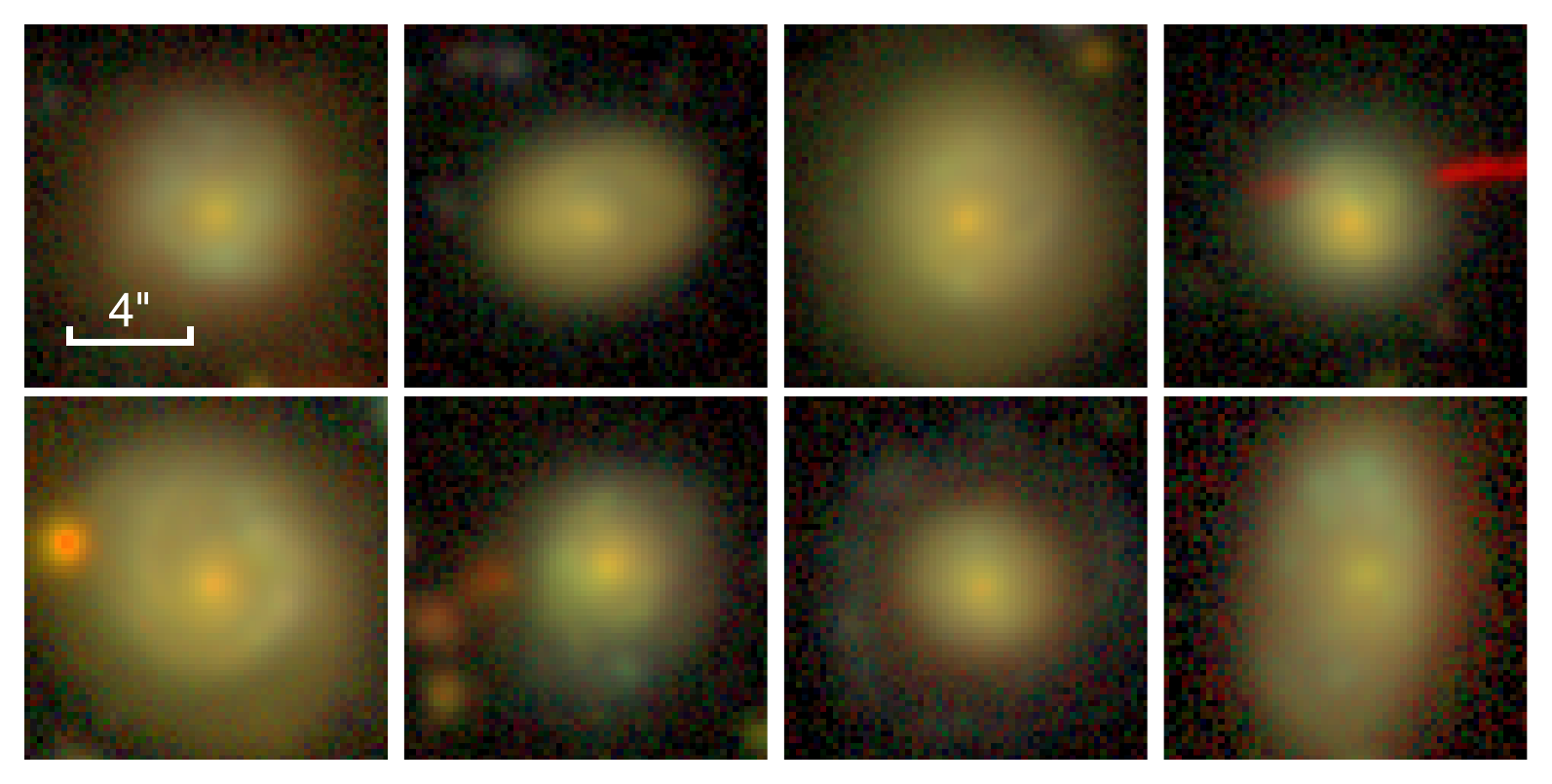}\label{fig:groups:c}}
    \subfigure[Cluster 127: \textit{Clumpy discs.}]{\includegraphics[width=0.43\textwidth]{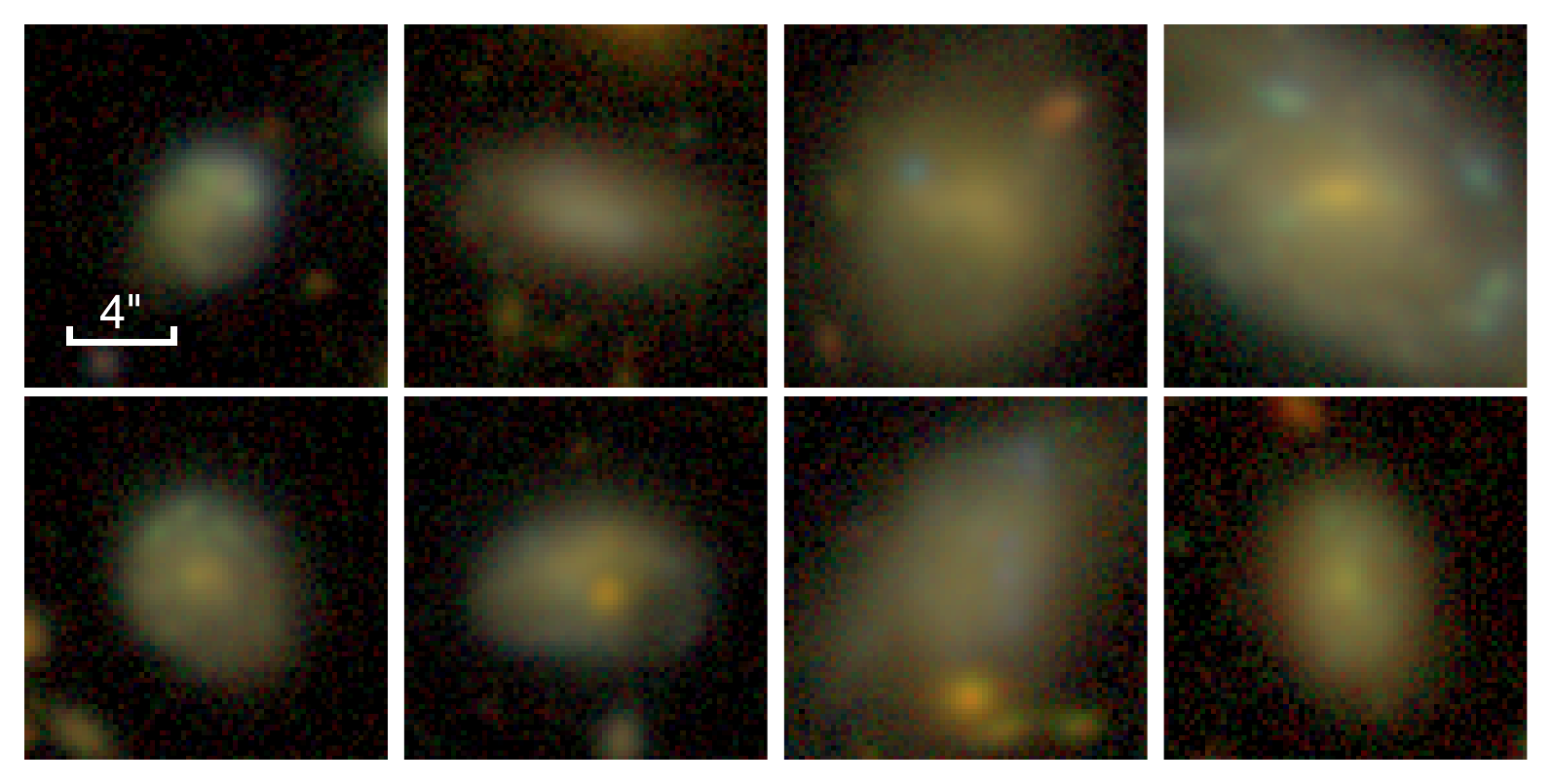}\label{fig:groups:e}}
    \caption{Examples of interesting morphological clusters produced by the algorithm: (a) Sc/Sd galaxies (b) merging systems at high-redshift (c) disks with blue ring-like structures that might be the result of the recent accretion of gas-rich satellites (d) clumpy disks in the nearby Universe. Spatial scales are indicated by the white bar in the top-left panel of each cluster.}
    \label{fig:groups}
\end{figure}

Fig \ref{fig:groups} shows some individual morphological clusters identified by the algorithm. For example, cluster 10 contains galaxies identified as Sc/Sd Hubble types (Fig \ref{fig:groups:a}), cluster 14 is comprised of systems that appear to be high-redshift mergers (Fig \ref{fig:groups:b}), cluster 122 contains galaxies which show blue ring-like features indicative of the recent accretion of gas-rich satellites (Fig \ref{fig:groups:c}) and cluster 127 is composed of clumpy discs (Fig \ref{fig:groups:e}). As described in Section \ref{sec:release} below, the visual classifications of each morphological cluster, and the average properties of objects in these clusters, are presented in Appendix \ref{app:tables} and Table \ref{table:props}. 


\subsection{Star-galaxy separation}
\label{sec:star_separation}

\begin{figure}
    \centering
    \includegraphics[width=0.45\textwidth]{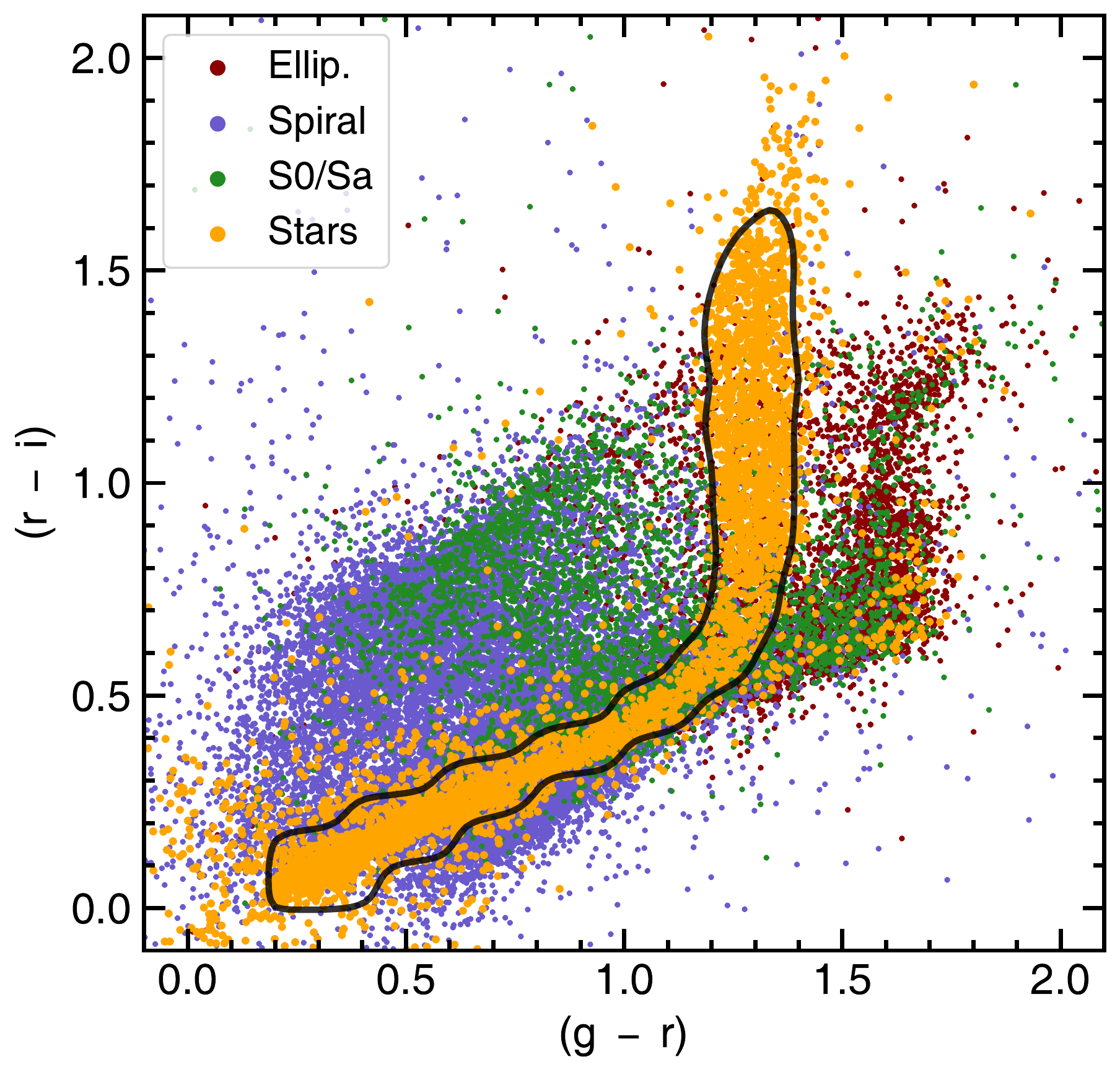}
    \caption{The positions of ellipticals (red), spirals (blue), S0/Sa galaxies (green) and stars (orange) in the rest-frame $g-r$ vs $r-i$ plane. The black contour contains 95 per cent of objects classified as not extended in the HSC-SSP catalogue.}
    \label{fig:stellar_locus}
\end{figure}

Figure \ref{fig:stellar_locus} presents a colour-colour diagram, showing the $g-r$ and $r-i$ colours for spirals (blue), S0/Sa galaxies (green), ellipticals (red) and stars (orange). The stellar locus is clearly delineated, occupying a distinct region of colour-colour space compared to spirals, ellipticals and S0/Sa galaxies. The objects that are morphologically identified as stars by the graph-clustering algorithm occupy the same region as objects that are identified as being not extended by the HSC pipeline (as defined in Section \ref{sec:classification}). The region of colour-colour space containing objects that are not extended is indicated by a black contour, which contains 95 per cent of all such objects in our sample.

It is worth noting that optical colours alone may not encode sufficient information to effectively separate stars from galaxies \citep{Fadely2012}. However, the algorithm employed here is able to distinguish between stars and resolved galaxies, even within the region where they share the same colours, because resolved galaxies and unresolved stars do not share the same (spatial) power spectra or distribution of patch types, and therefore, do not fall into the same morphological clusters. 


We note that the relatively simple method used in \citet{Eisenstein2001} (as described in Section \ref{sec:classification}) to determine extendedness is not always a good proxy for stellarity.
Although star-galaxy separation has traditionally used purely morphometric information to classify stars and galaxies in optical survey data \citep[e.g.][]{Kron1980, Eisenstein2001, Henrion2011}, new ground-based deep-wide surveys, which contain many more unresolved galaxies than stars at faint apparent magnitudes \citep{Fadely2012,Soumagnac2015}, represent an emerging challenge. Further work is therefore needed in order to determine whether the algorithm can effectively distinguish faint, unresolved galaxies from stars in very deep images. We note, however, that this does not affect the analysis in this study, since we are focused on bright objects. 


\subsection{Completeness of the UML-classified galaxy sample}

\begin{figure}
\centering
\includegraphics[width=0.45\textwidth]{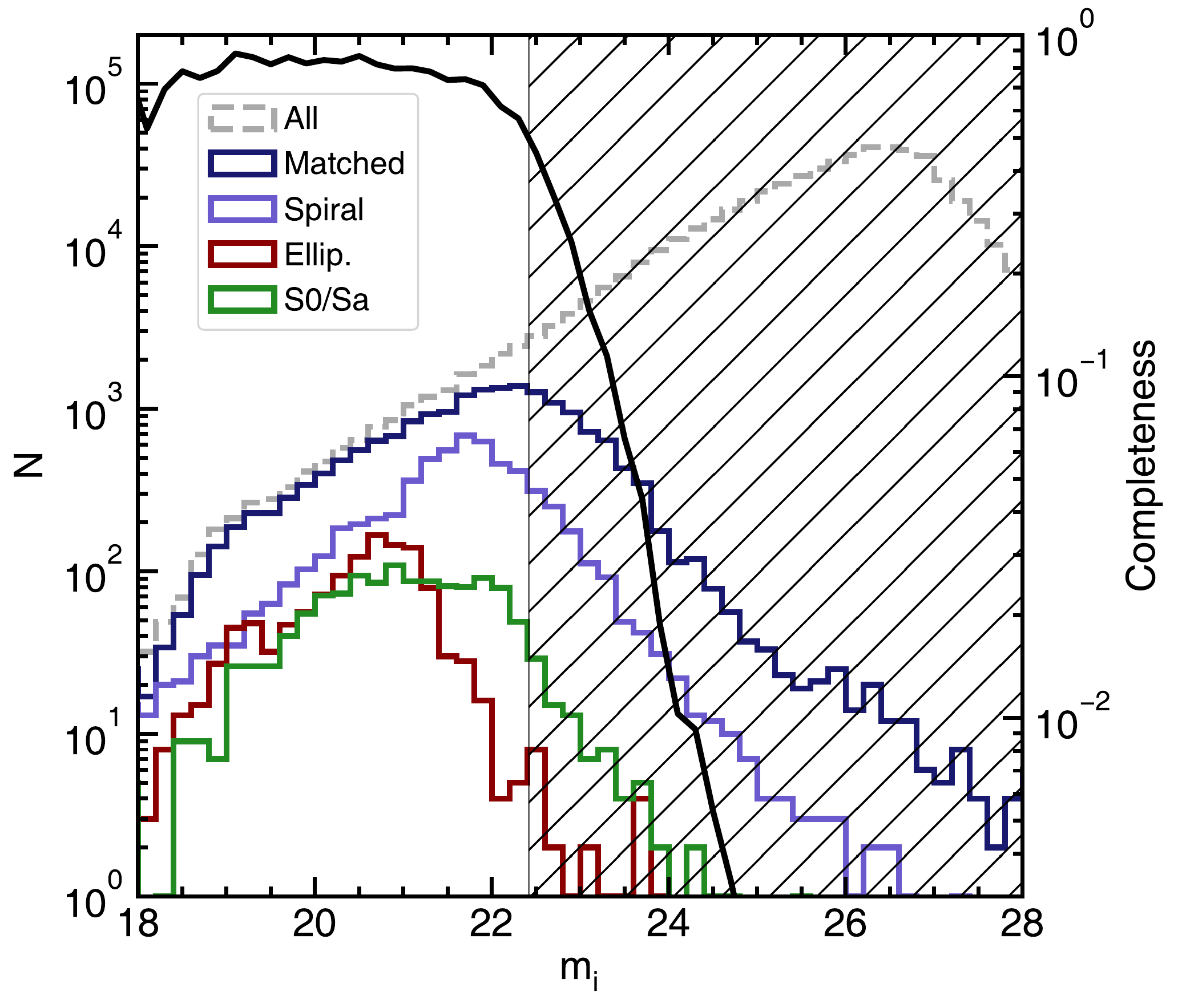}
\caption{The $i$-band apparent-magnitude distribution in the HSC-SSP Ultradeep catalogue (dashed grey histogram) and the distribution of the subset of these sources that has been matched and classified by the graph-clustering algorithm (solid dark-blue histogram). Light-blue, red and green histograms indicate the distribution of objects that are identified as spirals, ellipticals or S0/Sa galaxies respectively. The black solid line indicates the completeness as a function of apparent magnitude (with values indicated on the $y$-axis) and the grey hatching indicates the region where the completeness falls below 50 per cent.}     \label{fig:sb_test}
\end{figure}

In Fig \ref{fig:sb_test}, we compare the distribution of $i$-band magnitudes from the HSC-SSP DR1 Ultradeep survey (grey) and the distribution for objects that are large enough to be classified by the graph-clustering algorithm and then successfully matched to the HSC-SSP DR1  Ultradeep catalogue (dark blue). The black line indicates the completeness, i.e. the fraction of all galaxies in each magnitude bin that can be classified by the algorithm and then matched to the HSC-SSP. The completeness values are indicated by the right-hand $y$-axis. 

While the completeness of the full sample only begins to decline significantly around $m_{i}>27$~mag, a magnitude cut of $m_{i}<22.5$~mag ensures that a majority (i.e. more than 50 per cent) of objects in the Ultradeep survey have large enough sizes for robust morphological classification using the graph-clustering algorithm. We note that this cut appears to vary as a function of galaxy morphology, as demonstrated by the brighter limiting magnitudes for some morphologies, particularly for ellipticals. This is likely the result of different average projected sizes, with ellipticals being typically more compact than spiral galaxies that have similar magnitudes, particularly at low luminosities \citep[e.g.][]{Lange2015}. In the subsequent figures in this section, we consider galaxies brighter than this $m_{i}=22.5$~mag threshold. We also show that the size criterion imposed by the algorithm does not produce biased galaxy populations as a result of the classification and matching procedure. 


\begin{figure}
    \centering
    \subfigure[]{\includegraphics[width=0.45\textwidth]{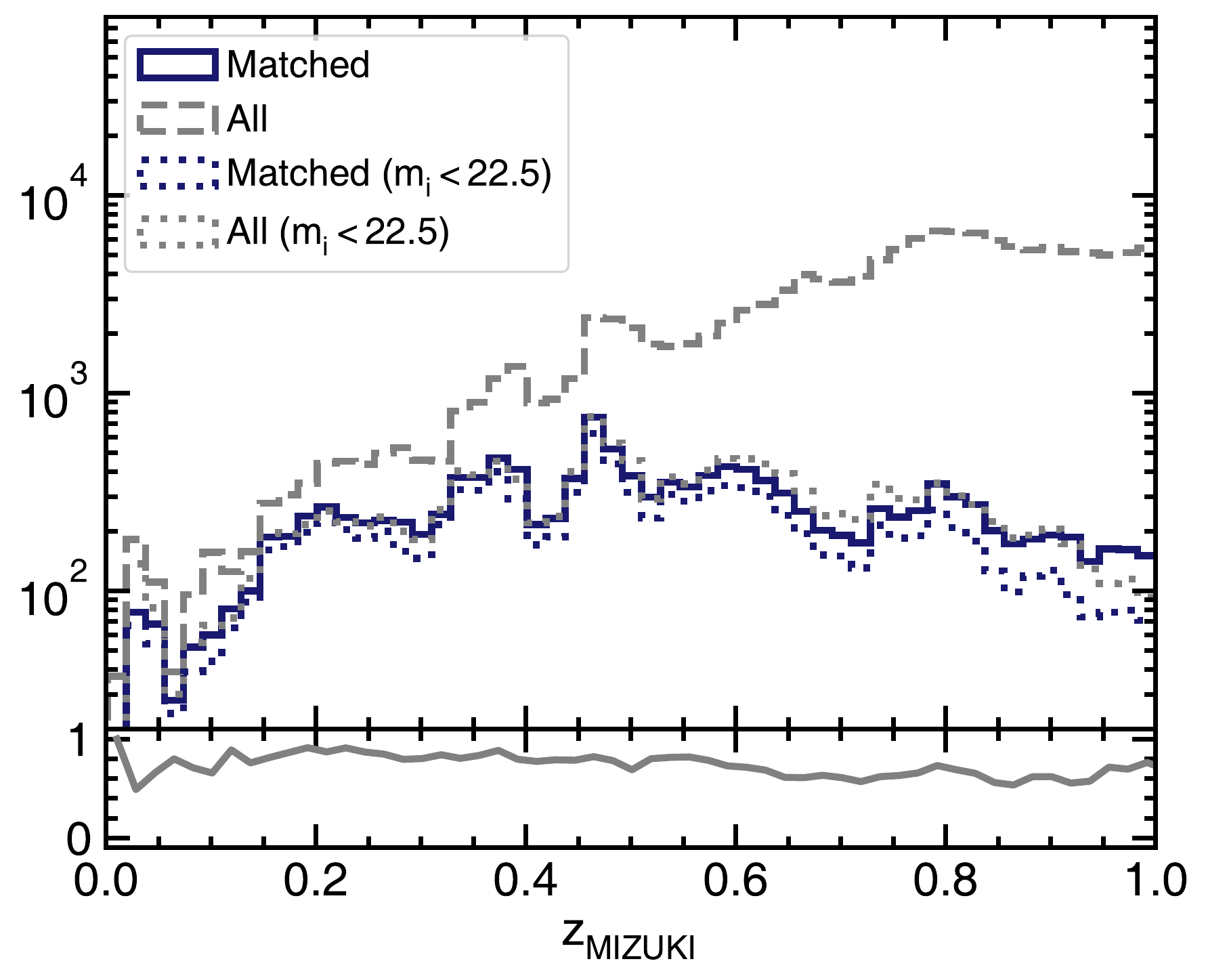}\label{fig:tests:a}}
    \subfigure[]{\includegraphics[width=0.45\textwidth]{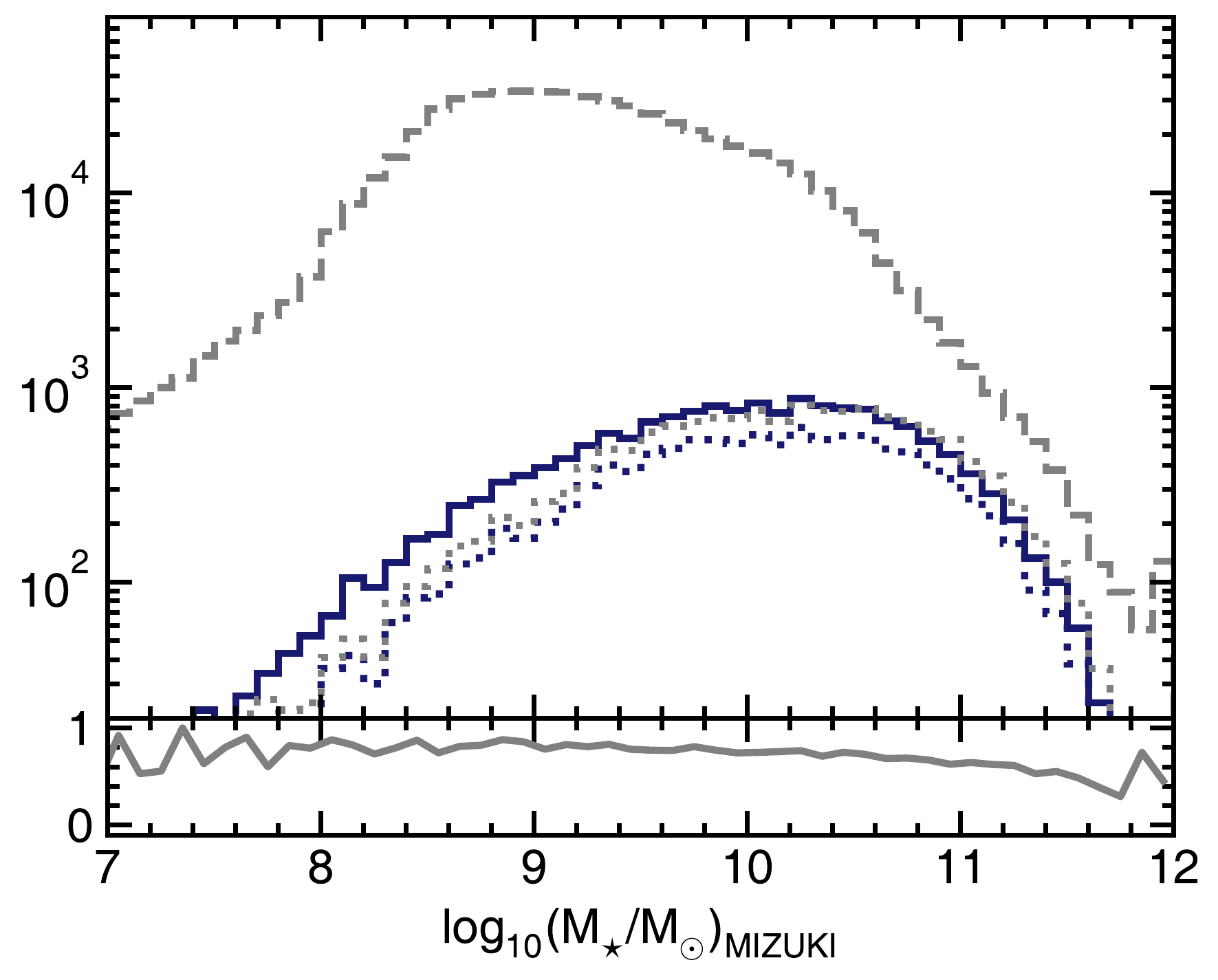}\label{fig:tests:b}}
    \subfigure[]{\includegraphics[width=0.45\textwidth]{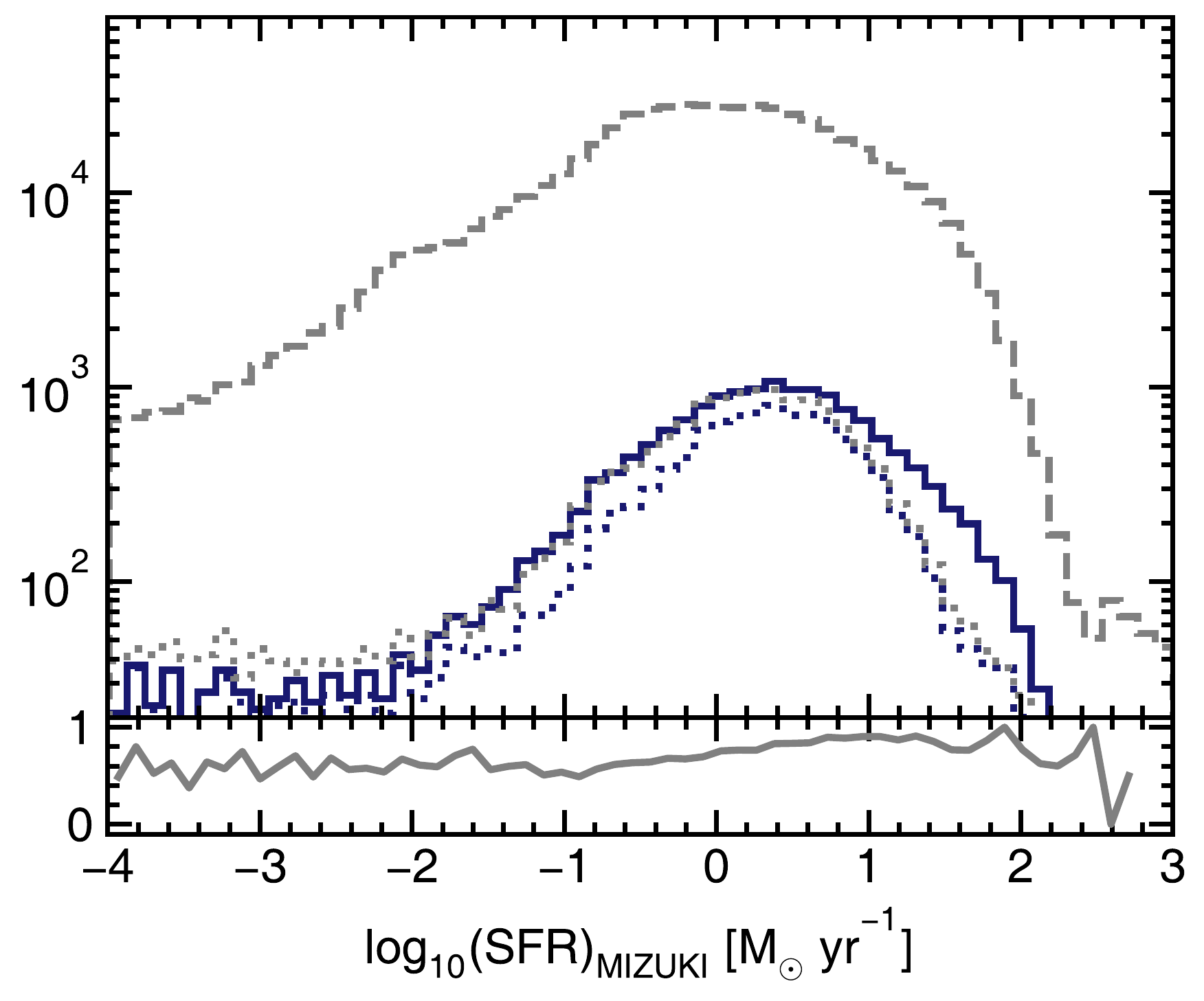}\label{fig:tests:c}}
    \caption{Distributions of galaxy properties (\textbf{(a)} \textsc{mizuki} photometric redshifts, \textbf{(b)}: stellar masses, \textbf{(c)}: SFRs) in the full HSC-SSP Ultradeep catalogue (grey) and for the subset of galaxies that has been matched and classified by the algorithm (blue). Dotted blue and grey lines show the distributions for galaxies with $m_{i} < 22.5$~mag. \textbf{Bottom}: The fraction of matched and classified objects with $m_{i} < 22.5$~mag as a function of photometric redshift.} 
    \label{fig:tests}
\end{figure}


Fig \ref{fig:tests:a} shows the distribution of photometric redshifts derived using the \textsc{mizuki} code, for the full Ultradeep catalogue (grey) and the matched catalogue from the algorithm (dark blue). Dotted grey and dark blue lines show the same for all galaxies with $i$-band absolute magnitudes brighter than 22.5~mag for the full and matched samples respectively. The lower panel shows the fraction of galaxies with $m_{i} < 22.5$~mag that are matched as a function of photometric redshift. As might be expected, the matched sample, the magnitude limited matched sample and the full magnitude limited sample all share similar distributions, but the matched sample falls off more quickly compared to the full sample, as their projected sizes increase with redshift, making more objects unclassifiable. 



Figs \ref{fig:tests:b} and \ref{fig:tests:c} show the corresponding analyses for stellar masses and SFRs respectively. Again, the histograms show the distributions of stellar masses and SFRs for the full HSC-SSP Ultradeep catalogue (grey), the distribution for matched objects only (dark blue) and the full and matched distributions for $m_{i}<22.5$~mag (shown using grey and blue dotted lines). The lower panels again show the fraction of galaxies with $m_{i} < 22.5$~mag that are matched as a function of photometric redshift. While the redshift distributions of objects is influenced by the size cut, for objects brighter than the magnitude cut, the full and matched samples have very similar distributions of physical properties. This indicates that the size cut does not introduce any bias in such galaxy properties (e.g. high SFRs), so that a comparison of average properties as a function of redshift is possible. 


\subsection{Released data products}
\label{sec:release}

The released data products for the HSC-SSP DR1 are contained in two tables. The first table comprises a list of morphological clusters with their associated visual classifications and median values of key galaxy properties within the cluster (surface-brightness, stellar mass, specific SFR, rest-frame $(g-r)$ colour and absolute $r$-band magnitude). This table is presented in its entirety in Appendix \ref{app:tables}. The second table is a list of individual HSC-SSP galaxies with their associated morphological cluster number and useful ancillary information, including their coordinates, HSC-SSP DR1 ID, extendedness, size in pixels and silhouette score. Note that some morphological clusters can have some contamination from stars. Users should discard objects which are classified as not extended and which are, therefore, likely to be stars. The first ten rows of this table is presented in Table \ref{tab:galaxy_properties} of Appendix \ref{app:tables}.

Both tables are available at the following URL: \url{https://github.com/garrethmartin/HSC_UML}.

\subsubsection{HSC-SSP DR2 release}

More comprehensive data products will be made available for the newer HSC-SSP DR2 UDEEP/DEEP data release at the same URL (\url{https://github.com/garrethmartin/HSC_UML}). We will provide multiple catalogues for different numbers of clusters (values of $k$ from 2 - 10, then increasing in increments of 10 up to 200). We will give average silhouette scores for each catalogue as a whole, in order to allow users to select the optimal number of clusters if desired, as well as provide diagnostic and silhouette plots for each cluster. We will provide the feature vectors for each galaxy with code for performing searches on these feature vectors to find similar objects \citep[i.e. a `similarity search',][]{Hocking2018} as well as code to run the entire algorithm on other data if desired.

\section{Galaxy properties as a function of morphological type}
\label{sec:analysis}

In this section, we explore the robustness of the morphological classifications produced by our algorithm. We study the distributions of key galaxy properties (e.g. stellar masses, star formation rates, rest-frame colours) as a function of morphological type, as a test of the veracity of our classifications. We demonstrate that the distributions of such galaxy properties in well-known morphological groups follow expected trends from studies using traditional visual morphological classification methods \citep[e.g][]{Menanteau2006,Kelvin2014,Khim2015,Willett2017}.


\subsection{Sample selection and methodology}


\subsubsection{Redshift binning}

\begin{figure}
\centering
\includegraphics[width=0.45\textwidth]{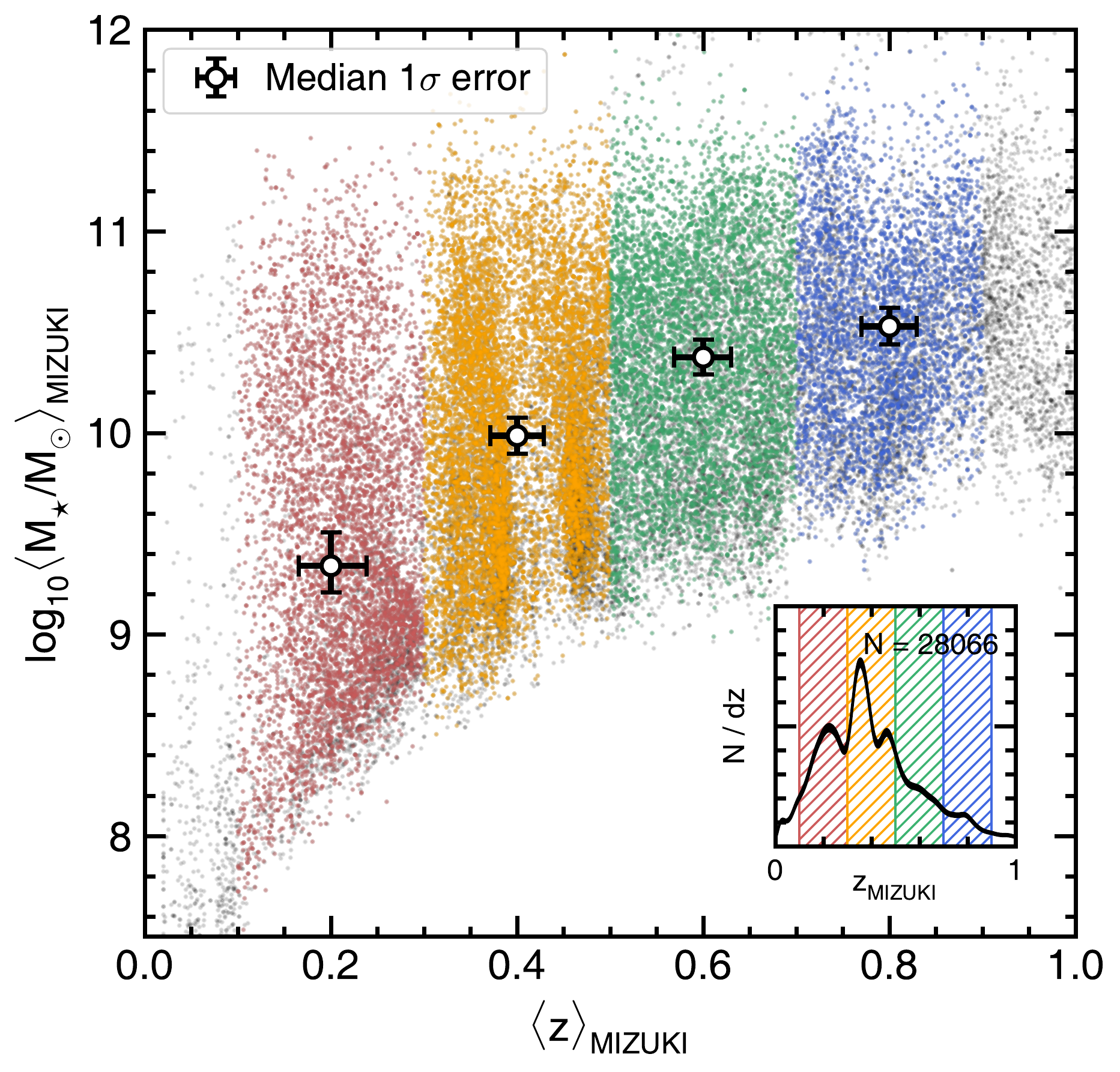}
\caption{\textbf{Main panel:} The redshift and stellar mass distributions of our sample for $m_{i}<22.5$ mag. The $x$- and $y$-axes indicate the median redshift and stellar mass respectively, derived from the \textsc{mizuki} fits to the optical SED of each object. Error bars indicate corresponding $1\sigma$ errors on the medians in each of the redshift bins. \textbf{Inset:} the \textit{unweighted} redshift distribution of our sample. The thickness of the line indicates the $2\sigma$ confidence interval, calculated using 10,000 draws from the redshift probability distribution of each galaxy, assuming a two-sided Gaussian error around the median value. Red, orange and green hatched regions indicate the three redshift bins used.}
    \label{fig:sample_z}
\end{figure}

We first bin our galaxies into three redshift ranges: $0.3<z<0.5$, $0.5<z<0.7$ and $0.7<z<0.9$. A minimum redshift of $z=0.3$ is chosen to ensure that the 3.5 deg$^{2}$ Ultradeep footprint encompasses a cosmological co-moving area greater than $85~\mathrm{Mpc}\times85~\mathrm{Mpc}$ in the lowest redshift bin, so that our galaxy populations are large enough to be statistically representative, and unlikely to be significantly biased by large-scale structure. For completeness, we do additionally consider a lower redshift bin ($0.1 < z < 0.3$) when studying stellar mass functions in Section \ref{sec:stellar_mass}, but this bin is likely to be strongly affected by cosmic variance. 

Fig \ref{fig:sample_z} presents a scatter plot showing the distribution of median stellar masses of individual galaxies as a function of their median redshifts. Points are colour coded by their redshift bin. Open circles with error bars indicate the central redshift and median stellar mass of each redshift bin and the $1\sigma$ error of these quantities in each bin. The inset shows the redshift distribution of galaxies. The thickness of the line indicates the $2\sigma$ confidence interval, derived using 10,000 draws from the redshift probability distribution, which assumes a two-sided Normal error around the median redshift, with a standard deviation equal to the upper and lower redshift errors. 


\subsubsection{$1/V_{max}$ weighting and simulation of uncertainties}

In order to correct for Malmquist bias \citep{Malmquist1922}, we weight galaxy counts using $1/V_{max}$, the inverse of the maximum volume in which it would be possible to detect an object of a given luminosity \citep[e.g.][]{Schmidt1968, Weigel2016}. We do this by first making 10,000 random draws from the redshift probability distribution for each object. We assume that the probability density function (PDF) follows a two-sided Normal distribution, with a central value equal to the median \textsc{mizuki} redshift, $\langle z \rangle$, but with different standard deviations ($\sigma_{upper}$ and $\sigma_{lower}$) on either side of the central value. Each redshift, $z_{draw}$, is therefore drawn from the following distribution:

\begin{equation}
    z_{draw} \sim 
    \begin{cases}
      \mathcal{N}(z\ |\ \langle z \rangle, \sigma^{2}_{lower}) \ \textrm{if} \ z\le \langle z \rangle\\
      \mathcal{N}(z\ |\ \langle z \rangle, \sigma^{2}_{upper})\times(\sigma_{lower}/\sigma_{upper}) \ \textrm{if} \ z>\langle z \rangle
    \end{cases}
\end{equation}

\noindent where $\mathcal{N}(z\ |\ \mu, \sigma^{2})$ is a Normal distribution with a central value equal to the median \textsc{mizuki} redshift, $\langle z \rangle$, and a variance of $\sigma^{2}$. $\sigma_{upper}$ is the 84th percentile of the redshift PDF, and $\sigma_{lower}$ is the 16th percentile of the redshift PDF. The factor of $\sigma_{lower}/\sigma_{upper}$ ensures that the distribution remains continuous.

In the \textsc{mizuki} fitting, the dominant source of uncertainty in the inferred stellar mass and absolute magnitude is the luminosity distance, rather than the model template weights or dust attenuation. The equivalent absolute magnitude, $M_{i, draw}$, and stellar mass, $M_{\star, draw}$, at a given redshift can therefore be well approximated by only varying their values by the square of the ratio of the luminosity distance, $D_{L}(z)$, at the redshift of the draw and the median redshift. We can therefore calculate the new stellar mass for each drawn redshift as follows:

\begin{equation}
M_{\star, draw} \approx \langle M_{\star}\rangle \big[ D_{L}(z_{draw}) / D_{L}(\langle z \rangle) \big]^{2}
\end{equation}

\noindent and similarly for the absolute magnitude:

\begin{equation}
M_{i, draw} \approx \langle M_{i} \rangle \big[ D_{L}(z_{draw}) / D_{L}(\langle z \rangle) \big]^{2}.
\end{equation}

\noindent We then find the maximum redshift, $z_{max}$ at which an object with absolute magnitude $M_{i,draw}$ will fall below the detection limit (the redshift where the distance modulus, $\mu$, is equal to $m_{lim, i} - M_{i,draw}$) and thus obtain $V_{max}$, which is proportional to the co-moving volume out to $z_{max}$.

\begin{equation}
    V_{max} \propto D_{c}(z_{max})^{3}, 
\end{equation}

\noindent where $D_{c}(z)$ is the comoving distance at $z$.

Note that the minimum size (15 pixels) that we impose influences the limiting apparent magnitude. Since the average size of objects at a given magnitude varies between morphological types, we use different values of $m_{lim,i}$ when calculating $z_{max}$, corresponding to the limiting magnitude found for the morphological type in question (e.g. as in Fig \ref{fig:sb_test}). Since we are primarily interested in the relative distribution of galaxy properties between morphological types, rather than the exact normalisation of the number density, we do not take into account the area of the survey when calculating $V_{max}$.

Following the method of weighting and simulating uncertainties described above, we take 10,000 draws from the redshift distribution for individual objects in each morphological type. For each of the 10,000 draws, we calculate new stellar masses and $i$ band absolute magnitudes and thus the value of $V_{max}$ for each galaxy. After binning our sample into four redshift bins, based on the draws from the redshift distribution (with central redshifts of 0.2 0.4, 0.6 and 0.8), we use $1/V_{max}$ weighted univariate Gaussian kernel density estimation \citep[e.g.][]{Klein2006} with a kernel bandwidth of 0.1~dex to produce a galaxy stellar mass function for each redshift bin. We use the median value and 1 and 2 $\sigma$ dispersions (defined by the central 68 and 95 per cent of values around the median) to characterise the galaxy stellar mass function and its uncertainty for each morphological type. 


\subsection{Stellar mass distributions as a function of morphological type}
\label{sec:stellar_mass}

\begin{figure}
    \centering
    \includegraphics[width=0.50\textwidth]{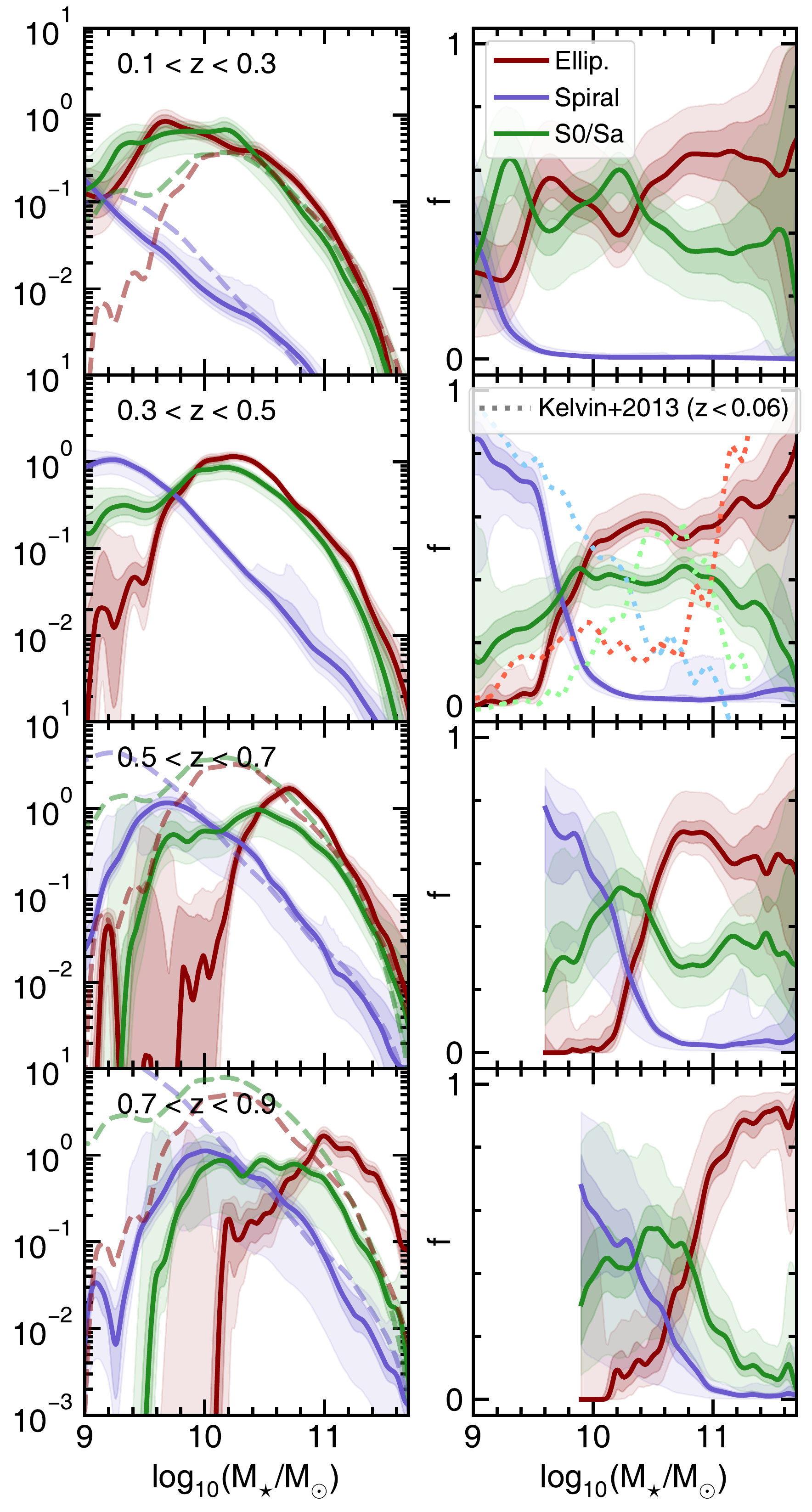}
    \caption{\textbf{Left}: Galaxy stellar mass functions for spirals (blue), S0/Sa galaxies (green) and ellipticals (red) in four redshift bins with arbitrary normalisation. Light and dark coloured regions indicate the $1\sigma$ and $2\sigma$ confidence intervals respectively, based on 10,000 draws from the redshift distribution of each galaxy. To enable comparison between the stellar mass functions at various redshifts, the pale dashed lines show the stellar mass function for the $0.3<z<0.5$ bin, normalised to the mass function in each redshift bin. \textbf{Right}: The evolution of the spiral, S0/Sa and elliptical fractions between $z=0.1$ and $z=0.9$. Blue, green and red lines show the fraction of galaxies that are spirals, S0/Sa and ellipticals, calculated from the galaxy stellar mass functions on the left. Light and dark coloured regions indicate the $1\sigma$ and $2\sigma$ confidence intervals respectively, based on 10,000 draws from the redshift distribution of each galaxy. Fractions are only plotted up to the point where the stellar mass function remains complete. Lighter dotted lines in the second to top panel indicate the elliptical, S0/Sa and spiral fractions from \citet[][Fig 3]{Kelvin2014} at $z<0.06$.}
    \label{fig:elliptical_fraction}
\end{figure}


Fig \ref{fig:elliptical_fraction} shows the evolution of the galaxy stellar mass function (left-hand column) and the evolution of the morphological fractions (right-hand column) as a function of redshift. We include a redshift bin in the range $0.1<z<0.3$ for completeness, however we avoid drawing any conclusions at these epochs, as the volume of this subsample is not large enough to be statistically representative, since it is likely to be strongly affected by cosmic variance. The light dotted lines in the second to top panel of the right hand column indicate the elliptical, S0/Sa and spiral fractions from \citet[][Fig 3]{Kelvin2014} at $z<0.06$. We note that the comparison is not perfect due to different definitions of morphological type, redshift range as well as cosmic variance. We attempt to create an equivalent definition for spirals by combining the LBS (little blue spheroid), Sab-Scd and Sd-Irr definitions from \citet[][Fig 3]{Kelvin2014}. The inclusion of irregular galaxies in \citep{Kelvin2014} may also be responsible for some of the discrepancy. However, regardless of any discrepancy, the general trend in spiral, elliptical and S0/Sa fractions remains the same, in agreement with other work \citep[e.g.][]{Conselice2008,Vulcani2011}

As shown in previous work \citep[e.g.][]{Conselice2008,Ilbert2010,Conselice2014}, there is a general trend for elliptical and S0/Sa fractions at a given stellar mass to increase towards lower redshifts. These systems increasingly dominate the number density at high stellar masses towards the present day \citep[e.g.][]{Wilman2012, Kelvin2014}, as spiral galaxies are quenched to form S0/Sa systems and/or undergo morphological transformation via mergers to form ellipticals. In the highest redshift bin ($0.7<z<0.9$), ellipticals almost entirely dominate at masses greater than $10^{11}$M$_{\odot}$, whereas S0/Sa galaxies become more important in the same mass range towards lower redshifts.

While S0/Sa galaxies and ellipticals share similar mass functions, at least at lower redshifts, the dominance of ellipticals at high stellar mass in the early Universe indicates that a distinct, more gradual, evolutionary channel may be responsible for producing the S0/Sa populations. In particular, ellipticals likely form at epochs that predate those where the mechanisms that produce S0/Sa populations \citep[e.g.][]{Dressler1997,Cerulo2017,Oh2019} are most efficient. This is likely to be particularly true for the most massive ellipticals, which must have formed rapidly at high or intermediate redshift \citep[e.g.][]{Jaffe2011,Tomczak2014,Company2015b}. 

The high-mass end of the elliptical mass function does not evolve significantly over redshift and is already in place in the highest redshift bin. The S0/Sa mass function appears instead to be built up from lower-mass systems, indicating a different evolutionary channel from their elliptical counterparts. At all redshifts, S0/Sa type galaxies typically dominate at intermediate masses, between spirals and ellipticals \citep[e.g.][]{Vulcani2011,Kelvin2014}, with the peak of the S0/Sa fraction moving towards lower stellar masses at lower redshifts.  


\subsection{Star formation rates and rest-frame colours as a function of morphological type}

\begin{figure}
\centering
\includegraphics[height=0.8\textheight]{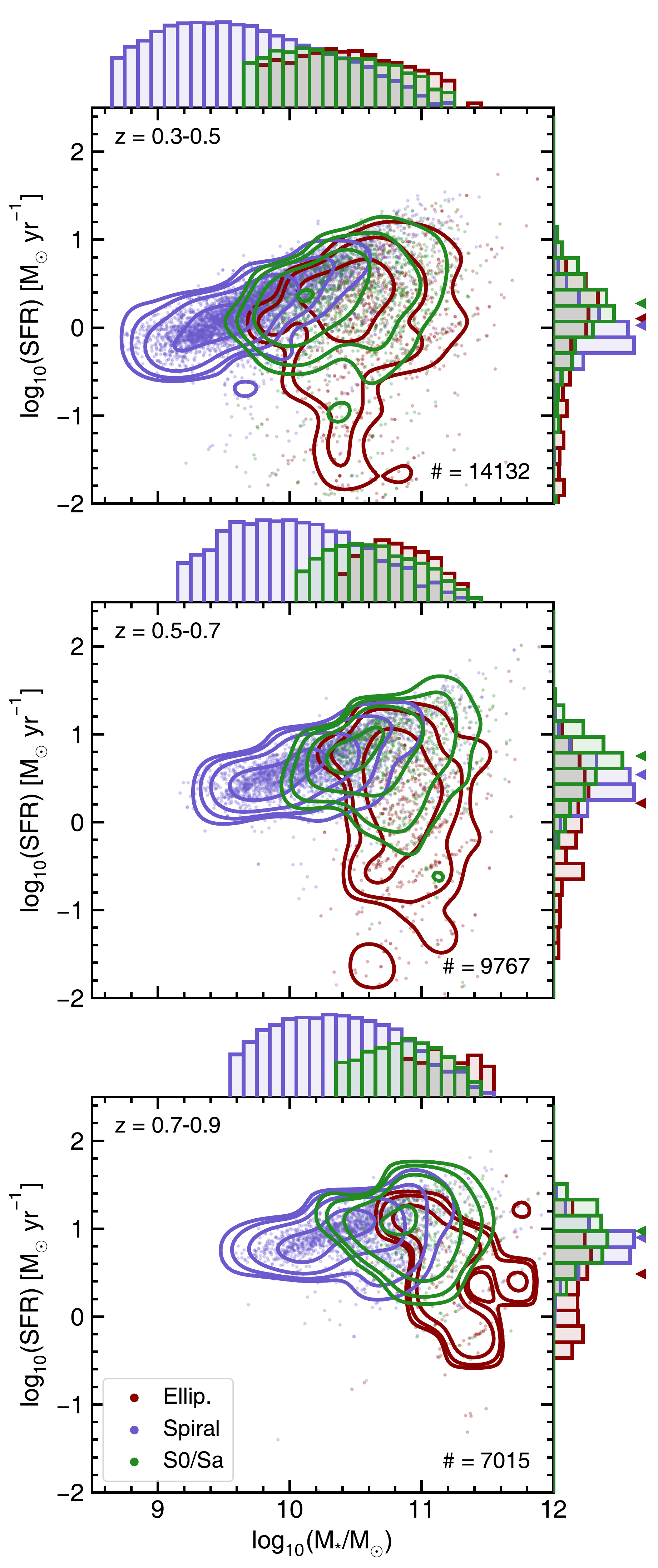}
\caption{Scatter plots with contours overlaid, showing the distribution of galaxies as a function of SFR and stellar mass, for galaxies classified as elliptical (red), S0/Sa (green) and spiral (blue). The dots show individual galaxies, while contours show the $1/V_{max}$ weighted density, with log$_{10}$ distributed levels. Each panel shows a different redshift range (using the \textsc{mizuki} derived photometric redshifts) which is indicated in the top-left corner. Histograms at the top and right-hand side of each panel show the distributions of stellar mass and SFRs respectively, for each morphological type. Coloured triangles indicate the $1/V_{max}$ weighted median SFRs for ellipticals, S0/Sa galaxies and spirals. The number in the bottom right corner of each panel indicates the total number of objects in each redshift bin.}
    \label{fig:dist_MS}
\end{figure}

\begin{figure}
\centering
\includegraphics[height=0.8\textheight]{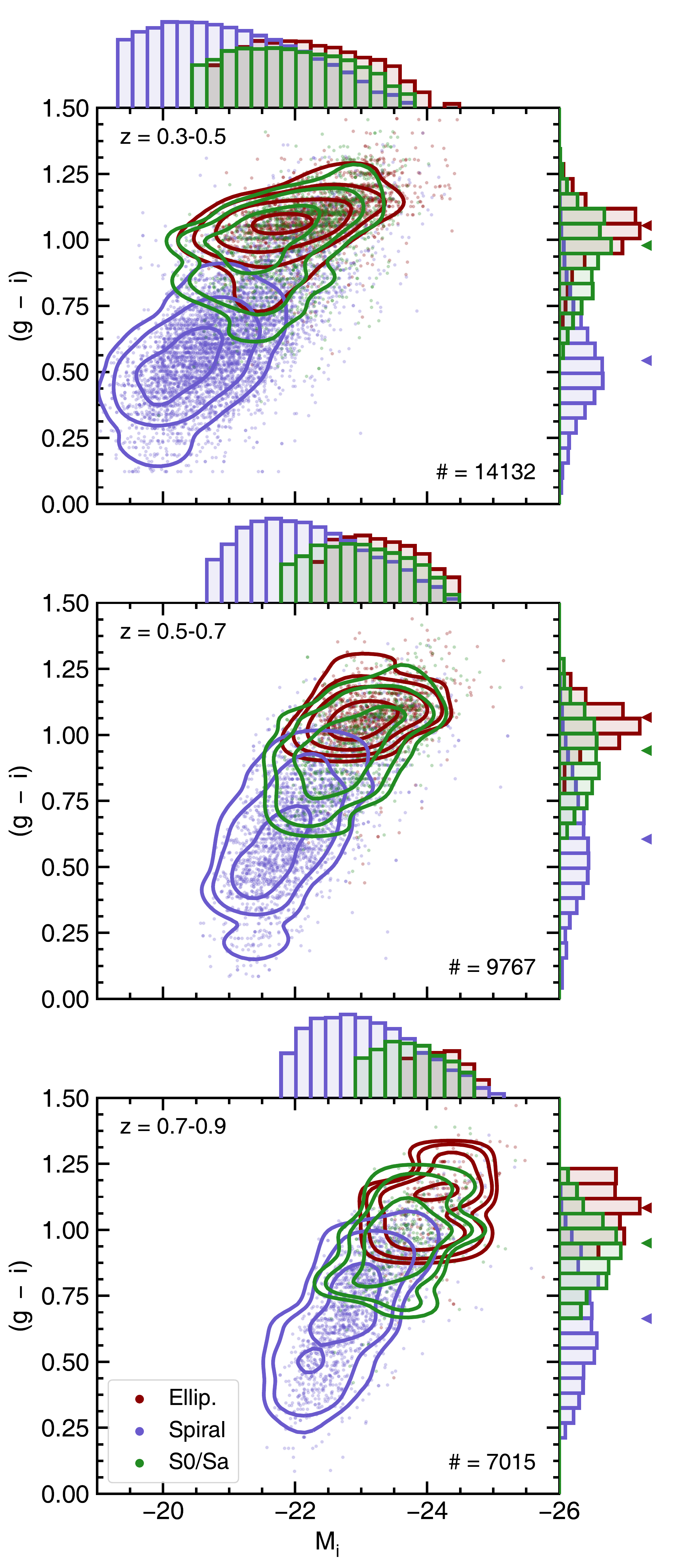}
\caption{Contour plots showing the distribution of galaxies as a function of $g-i$ colour and rest-frame $i$-band absolute magnitude, for galaxies that have been classified as elliptical (red), S0/Sa (green) and spiral (blue). Dots show individual galaxies, while contours show the $1/V_{max}$ weighted density with log$_{10}$ distributed levels. Each panel shows a different redshift range (using the \textsc{mizuki} derived photometric redshifts) indicated in the top left corner. Histograms at the top and right hand side of each panel show the distribution of rest-frame $i$-band magnitudes and $g-i$ colours respectively, for each morphological type. Coloured triangles indicate the $1/V_{max}$ weighted median $g-i$ colours for ellipticals, S0/Sa galaxies and spirals. The number in the bottom right corner of each panel indicates the total number of objects in each redshift bin.}
    \label{fig:dist_col_mag}
\end{figure}

Figs \ref{fig:dist_MS} and \ref{fig:dist_col_mag} show the star formation main sequence and the $M_{i}$ vs. rest-frame $g-i$ colour-magnitude diagram, for three redshift bins (with central redshifts of 0.4, 0.6 and 0.8). Contours show the density of objects weighted by $1/V_{max}$. Galaxies classified as spirals inhabit a well-defined main sequence (Fig \ref{fig:dist_MS}), while ellipticals dominate a cloud below this sequence. S0/Sa galaxies lie somewhere between these two populations. Many S0/Sa galaxies are not quenched and remain on the main locus of the star formation main sequence, with a small number lying further below. Similarly, the colour-magnitude diagram (Fig \ref{fig:dist_col_mag}) shows a clear bi-modality, with galaxies classified as ellipticals occupying the `red sequence' and galaxies classified as spirals occupying the `blue cloud' \citep[e.g][]{Baum1959,Visvanathan1981}. 
S0/Sa galaxies inhabit both parts of the diagram, but largely occupy the space in between the two distributions defined by the spiral and elliptical populations. 

The histograms above each panel in Fig \ref{fig:dist_MS} show the distribution of stellar masses for each morphological type. In agreement with other studies \citep[e.g.][]{Kelvin2014}, we find that the stellar mass function of S0/Sa galaxies is much closer to that of ellipticals than spirals. Spirals are much less massive, on average, than ellipticals and S0/Sa galaxies, while S0/Sa galaxies have marginally lower stellar masses than ellipticals. 

The histograms on the right-hand side of each panel in Fig \ref{fig:dist_MS} show the distributions of SFRs. Coloured arrow heads indicate the $1/V_{max}$ weighted median values \citep[e.g.][]{Edgeworth1888} of the SFRs in each population. While S0/Sa galaxies typically have SFRs that are comparable to spirals and higher than those found in ellipticals, they are typically more massive and therefore inhabit an intermediate range of values of specific SFRs. They remain redder and less star-forming than the majority of spirals, although the majority retain fairly high levels of star formation compared to ellipticals \citep[e.g.][]{Thronson1989,Pogge1993}.

The histograms above each panel in Fig \ref{fig:dist_col_mag} show the distributions of absolute $i$-band magnitudes for each morphological type, while the histograms to the right of each panel show distributions of rest-frame $g-i$ colours. Again, coloured arrow heads indicate the $1/V_{max}$ weighted median values for each population. Galaxies classified as ellipticals and spirals inhabit opposite ends of a bi-modal distribution in $g-i$ colour, with galaxies classified as S0/Sa typically lying between the two populations \citep[e.g][]{Wilman2012,Lopez2018}.

Given that different morphologies show some separation in integrated properties (e.g. stellar mass and SFR), it may be tempting, particularly when faced with the data volumes expected from future surveys, to use these properties as proxies for morphology. However, as previous studies have shown \citep[e.g.][]{Fadely2012,Vika2015}, spatial frequency information is essential for the robust morphological classification of both stars and galaxies (see also Section \ref{sec:star_separation}). We use our morphological classifications to explore this point in more detail. 

\begin{figure*}
    \centering
    \subfigure[]{\includegraphics[width=0.45\textwidth]{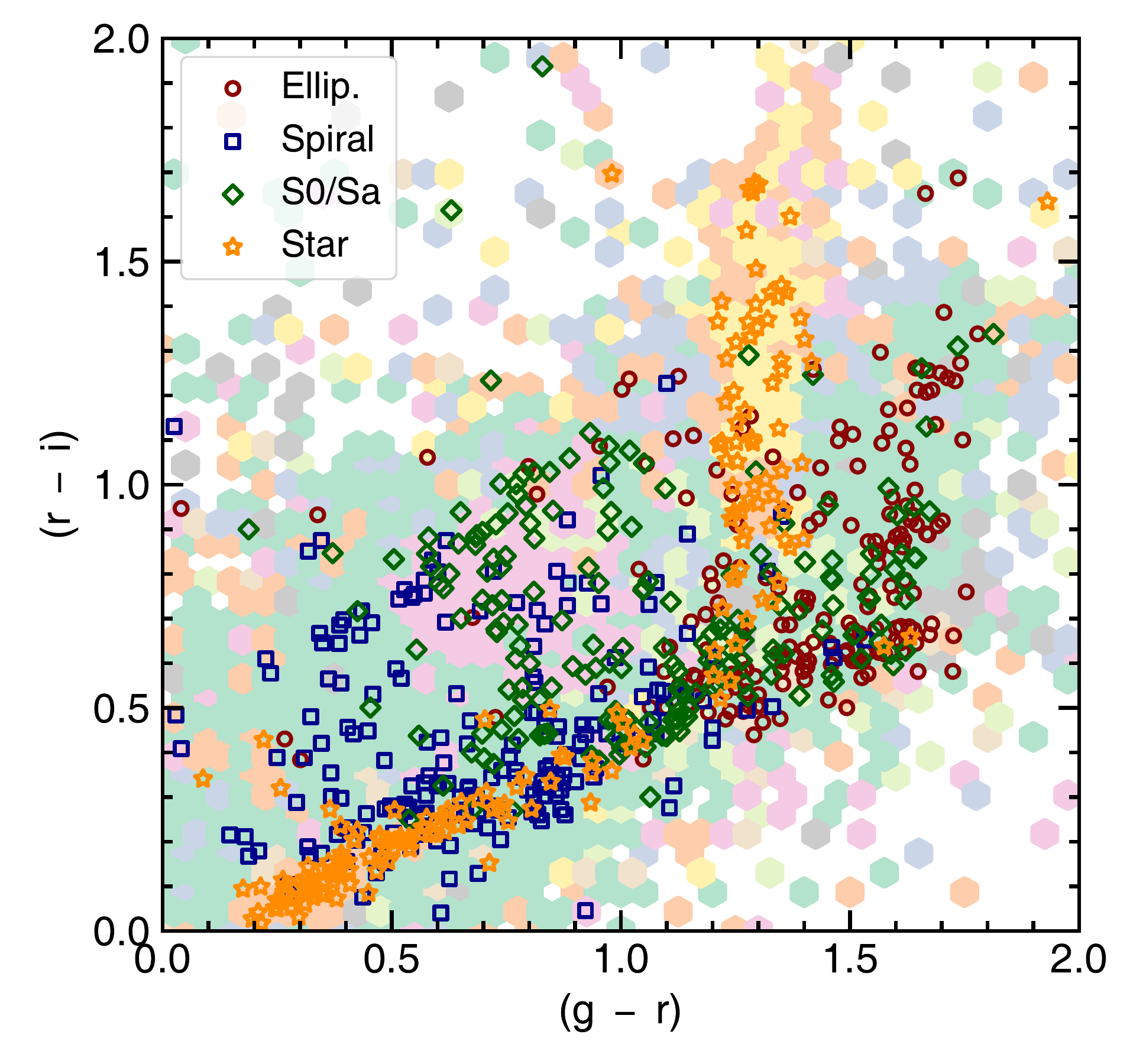}\label{fig:mode_groups:a}}
    \subfigure[]{\includegraphics[width=0.45\textwidth]{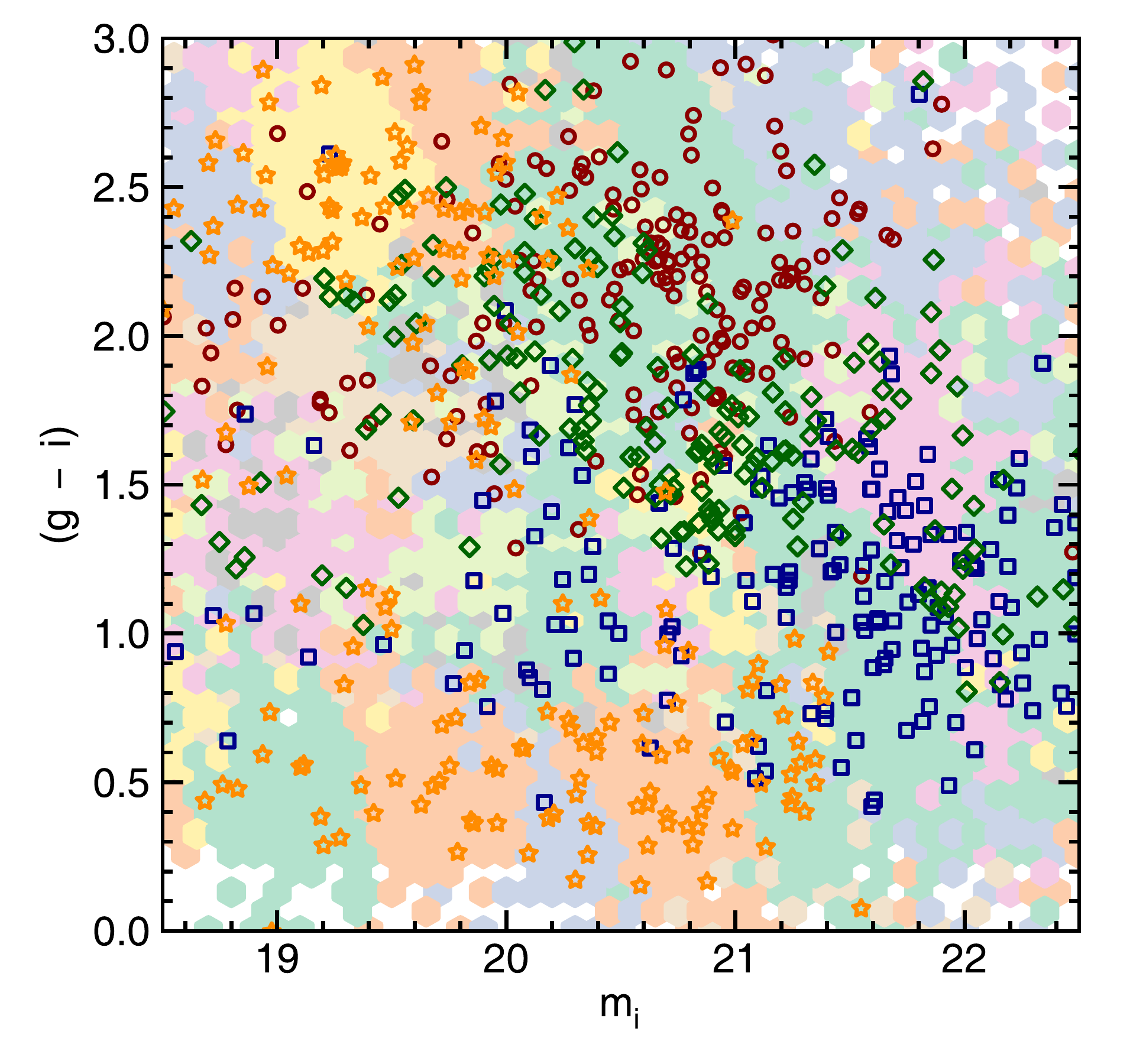}\label{fig:mode_groups:b}}
    \subfigure[]{\includegraphics[width=0.45\textwidth]{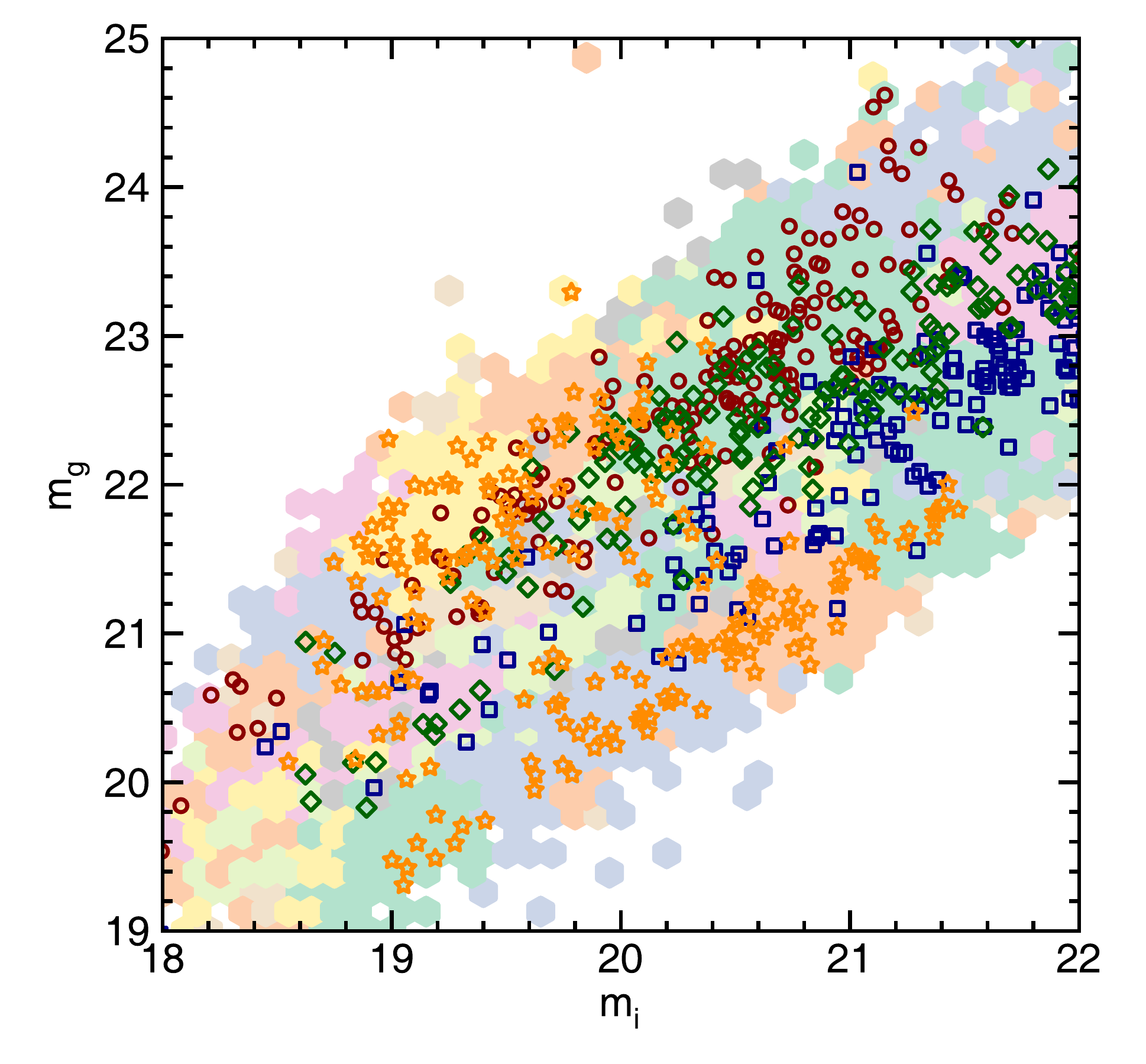}\label{fig:mode_groups:c}}
    \subfigure[]{\includegraphics[width=0.45\textwidth]{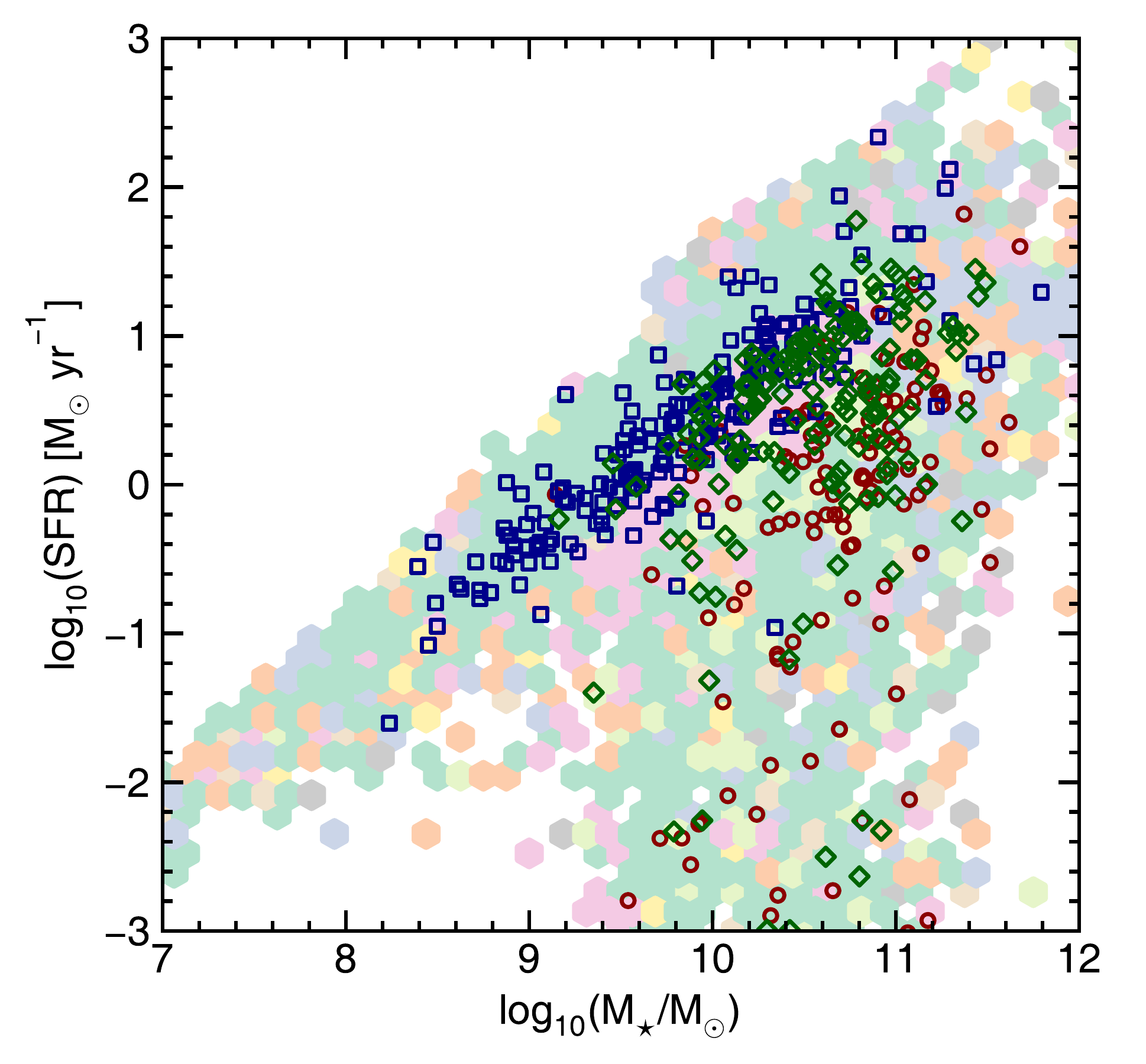}\label{fig:mode_groups:d}}
    \caption{Morphological clusters as a function of various parameters. Contiguous hexagonal bins with the same colour indicate regions in the parameter space which share the same dominant group. We show the most frequent morphological clusters in colour-colour \textbf{(a)}, colour-magnitude \textbf{(b)}, magnitude-magnitude \textbf{(c)} and stellar mass vs. star formation rate \textbf{(d)} space. Open red circles, blue squares, green diamonds and orange stars show the positions of a random sample of 200 ellipticals, spirals, S0/Sa galaxies and stars within each parameter space.}
    \label{fig:mode_groups}
\end{figure*}

Fig \ref{fig:mode_groups} shows the positions of a random selection of objects classified as ellipticals, spirals and S0/Sa galaxies within the colour-colour (Fig \ref{fig:mode_groups:a}), colour-magnitude (Fig \ref{fig:mode_groups:b}), magnitude-magnitude (Fig \ref{fig:mode_groups:c}) and stellar mass-SFR (Fig \ref{fig:mode_groups:d}) planes. Regions of contiguous colour in each plot indicate parts of the parameter space which are dominated by objects of a given morphological type i.e. the parameter space is colour-coded by the modal group in each hexagonal bin. It is clear that a significant fraction of objects of different morphology can fall into the same regions of parameter space, regardless of the exact plane being considered. Thus, a large degree of overlap exists in the integrated properties of S0/Sa galaxies, spirals and ellipticals, not only in colour-colour, colour-magnitude and magnitude-magnitude space, but also in physical properties like stellar mass and SFR. Such integrated properties \textit{alone} are therefore not sufficient to separate objects morphologically. The spatial information contained in the power spectrum of each patch type, as well as the spatial distribution of patch types across each object, are essential ingredients of accurate morphological classification. 

\section{Summary}
\label{sec:summary}

Morphology is a fundamental quantity that encodes the principal mechanisms that drive the evolution of individual galaxies. Essential for the full spectrum of galaxy-evolution studies, morphology is an important parameter for an array of topics in astrophysics, e.g. as a prior in photometric redshift pipelines and as contextual data in transient lightcurve classifications. A rich literature exists on morphological-classification techniques, with methods ranging from automated classification (e.g. via parametric and non-parametric reductions of galaxy images and machine-learning techniques) to direct visual classification by human classifiers, which is typically used to benchmark automated algorithms.   

Notwithstanding the array of techniques on offer, the forthcoming era of `Big Data' deep-wide surveys poses unique challenges for measuring galaxy morphologies. The sheer volume of data expected from surveys like LSST and Euclid makes visual classification intractable for such datasets (even via massively-distributed systems like Galaxy Zoo) and makes some degree of automation essential for this exercise. The short cadence of surveys like LSST presents an additional challenge, because repeatedly producing training sets, that are required for supervised machine-learning techniques, on short timescales may be impractical. 

Unsupervised machine-learning (UML) offers an attractive solution to these problems and an ideal route for the morphological classification of galaxies in next-generation surveys. An effective UML algorithm can autonomously compress an arbitrarily large galaxy population into a small set of morphological clusters whose members have similar morphology. If the number of clusters is small enough (e.g in the hundreds or less), then this makes it tractable to benchmark them using visual classification by individual researchers. The resultant classifications can thus combine both the speed of automation and the accuracy of visual classification. 

Here, we have employed such a UML algorithm, which automatically identifies distinct groups of galaxy types from survey pixel data, to separate galaxies in the HSC-SSP DR1 Ultradeep layer into 160 morphological clusters. This technique extracts sub-image patches from multi-band HSC data, each of which are transformed into a rotationally-invariant representation of a small region of the survey data, efficiently encoding colour, intensity and spatial frequency information. Utilising growing neural gas and hierarchical clustering algorithms, it then groups patches into a library of patch types, based on their similarity, and assembles feature vectors for each object, which describe the frequency of each patch type. A $k$-means algorithm is then used to separate objects into morphological clusters, based on the similarity of their feature vectors. 

We have visually inspected a representative sample of objects in each morphological cluster to classify them into three broad morphological types: elliptical galaxies, S0/Sa galaxies and spiral galaxies. We also provide finer morphological information e.g. the type of spiral morphology (Sb, Sc, Sd) and noteworthy colour or structural features (e.g. when spirals appear unusually red or show clumpy structure, or when elliptical galaxies appear unusually blue). To test the robustness of the classifications, we have shown that galaxies in different morphological classes reproduce known trends in key galaxy properties as a function of morphological types at $z<1$, e.g. stellar mass functions, rest-frame magnitudes and colours and the position of galaxies on the star formation main sequence. 

Our study demonstrates the potential of UML in the morphological analysis of forthcoming deep-wide surveys. The combination of initial UML-driven automation, followed by benchmarking via visual classification, is likely to become an optimal tool for the morphological analysis of surveys like LSST. While this study has focused on bright galaxies at $z<1$, it is worth noting that a significant fraction of objects, especially at low masses, inhabit the low-surface-brightness (LSB) Universe \citep[e.g.][]{Martin2019}. In forthcoming work, we will optimize the algorithm for the morphological classification of LSB galaxies and the detection of LSB structures, such as faint merger-induced tidal features, which will be routinely detectable in future surveys like those from the LSST. Furthermore, while our morphological classifications are limited to $z<1$, due to the ground-based nature of the HSC images, implementation of this UML algorithm on forthcoming higher-resolution data, e.g. from Euclid, will enable virtually all-sky morphological classification of galaxies out to high redshift.   


\section*{Acknowledgements}
We thank the anonymous referee, whose comments helped us improve the quality of this paper. We thank Dan Smith for many interesting discussions. GM and SR acknowledge support from the Science and Technology Facilities Council [ST/N504105/1]. SK acknowledges a Senior Research Fellowship from Worcester College Oxford. 

This work in based, in part, on data collected at the Subaru Telescope and retrieved from the HSC data archive system, which is operated by Subaru Telescope and Astronomy Data Center at National Astronomical Observatory of Japan. The Hyper Suprime-Cam (HSC) collaboration includes the astronomical communities of Japan and Taiwan, and Princeton University. The HSC instrumentation and software were developed by the National Astronomical Observatory of Japan (NAOJ), the Kavli Institute for the Physics and Mathematics of the Universe (Kavli IPMU), the University of Tokyo, the High Energy Accelerator Research Organization (KEK), the Academia Sinica Institute for Astronomy and Astrophysics in Taiwan (ASIAA), and Princeton University. Funding was contributed by the FIRST program from Japanese Cabinet Office, the Ministry of Education, Culture, Sports, Science and Technology (MEXT), the Japan Society for the Promotion of Science (JSPS), Japan Science and Technology Agency (JST), the Toray Science Foundation, NAOJ, Kavli IPMU, KEK, ASIAA, and Princeton University. This paper makes use of software developed for the Large Synoptic Survey Telescope. We thank the LSST Project for making their code available as free software at  \url{http://dm.lsst.org}.

The Pan-STARRS1 Surveys (PS1) have been made possible through contributions of the Institute for Astronomy, the University of Hawaii, the Pan-STARRS Project Office, the Max-Planck Society and its participating institutes, the Max Planck Institute for Astronomy, Heidelberg and the Max Planck Institute for Extraterrestrial Physics, Garching, The Johns Hopkins University, Durham University, the University of Edinburgh, Queen’s University Belfast, the Harvard-Smithsonian Center for Astrophysics, the Las Cumbres Observatory Global Telescope Network Incorporated, the National Central University of Taiwan, the Space Telescope Science Institute, the National Aeronautics and Space Administration under Grant No. NNX08AR22G issued through the Planetary Science Division of the NASA Science Mission Directorate, the National Science Foundation under Grant No. AST-1238877, the University of Maryland, and Eotvos Lorand University (ELTE) and the Los Alamos National Laboratory.

\bibliographystyle{mnras}
\bibliography{references}



\appendix


\section{Cross-matching detected objects to HSC-SSP DR1 centroids}
\label{app:cross_match}

\begin{figure}
    \centering
    \includegraphics[width=0.45\textwidth]{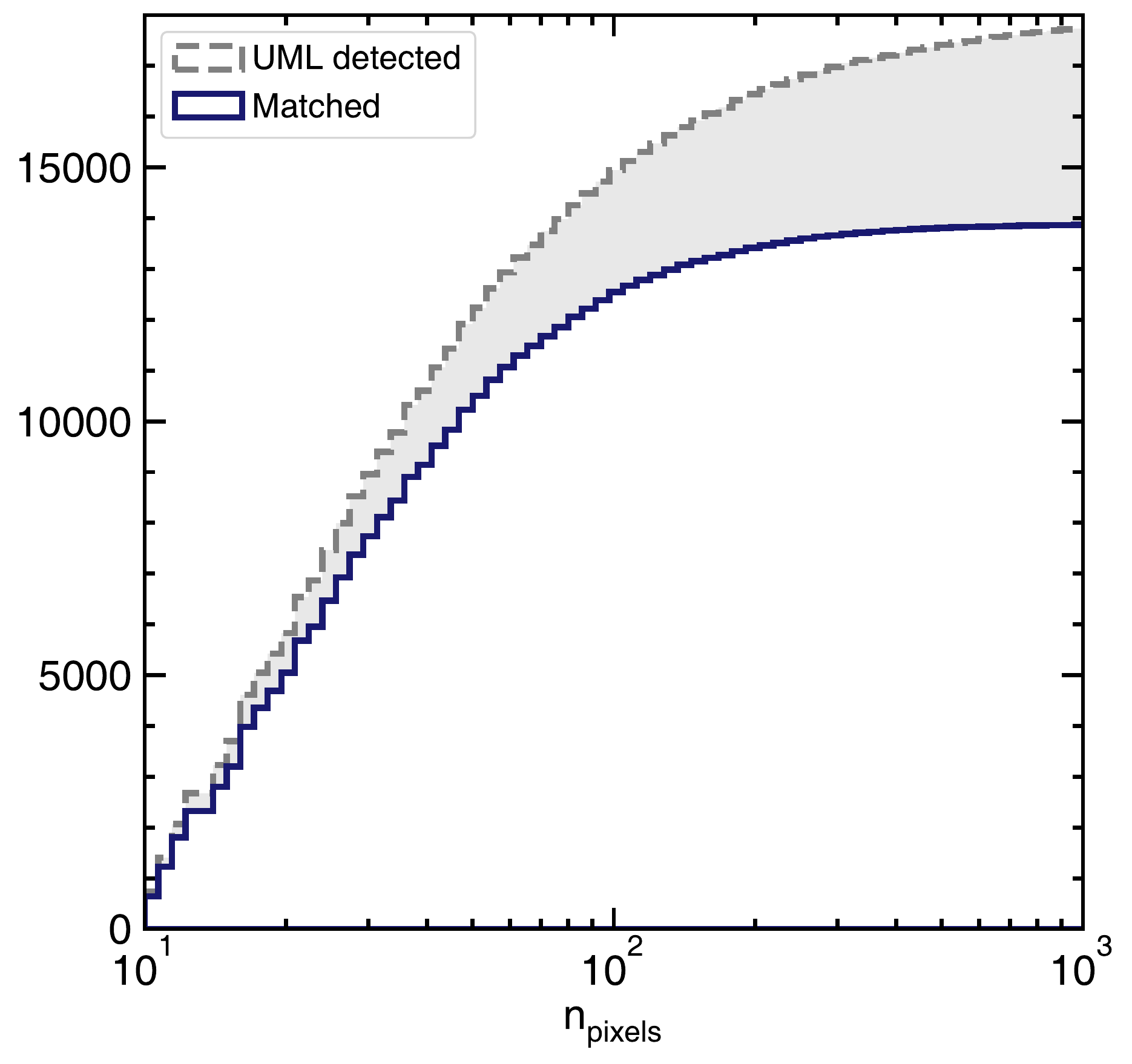}
    \caption{The grey dotted histogram indicates the cumulative number of objects within a single tract, with sizes larger than 10 pixels, that are detected by the algorithm as a function of their size in pixels. The blue solid histogram indicates the same for objects whose centroid is matched to within $0.8''$ of an object in the HSC-SSP DR1 catalogue. The shaded region shows the difference between the two histograms.}
    \label{fig:npix}
\end{figure}

Fig \ref{fig:npix} shows the cumulative frequency of objects detected by the algorithm within a single tract of HSC-SSP data, as a function of of their size. The size of each object is measured by the number of pixels it consists of, which is determined by the number of connected components above the sigma-clipping level in each object. The grey dashed histogram shows the cumulative number of objects larger than 10 pixels detected by the algorithm. The blue solid histogram indicates the cumulative number of these objects whose centroids can be matched to objects in the HSC-SSP DR1 catalogue within $0.8''$. Although the number of objects successfully matched is close to the total number of objects detected by the algorithm for small sizes ($n_{\mathrm{pixels}} \lesssim 50$), objects with larger sizes are significantly less likely to be matched.

The mismatch between centroids becomes a significant problem for large objects, with almost no objects with sizes larger than 200 pixels being matched. We note, however, that this mismatch does not present a significant problem for our analysis, as we consider only intermediate redshift ($z>0.3$) objects, which typically have small sizes. It would, however, be advantageous to select detected pixels from the object footprints taken directly from a catalogue that we hope to match to (i.e. in this case, from the stacked \texttt{calexp} images from HSC-SSP DR1) -- as discussed at the beginning of Section \ref{sec:analysis}. This is likely to yield more reliable cross-matching, especially for large, nearby objects and might be necessary for a perfect one-to-one matching. 



\section{Released data products - lists of morphological clusters and individual galaxy properties}
\label{app:tables}

In this Appendix, we present the tables that form the data release from this paper (\url{https://github.com/garrethmartin/HSC_UML}). Table \ref{tab:cluster_properties} describes the morphological clusters with their associated visual classifications and median values of key galaxy properties within the cluster (surface-brightness, stellar mass, specific SFR, rest-frame $(g-r)$ colour and absolute $r$-band magnitude). Table \ref{tab:galaxy_properties} (only the first ten rows are shown here) provides a list of individual HSC-SSP galaxies with their associated morphological cluster IDs and ancillary information. As noted above, users should discard objects which are classified as not extended, as they are likely to be stars.


\begin{table*}
\caption{Average quantities (and their $1\sigma$ dispersions) for objects in individual morphological clusters (the cluster ID is indicated by the first column, \#). Columns are as follows:  $(a)$ the number of \textit{matched} objects in the morphological cluster, $(b)$ the number of objects identified as extended by the HSC-SSP pipeline, $(c)$ median redshift, $(d)$ median surface brightness in mag~arcsec$^{-2}$, $(e)$ median stellar mass, $(f)$ median sSFR in M$_{\odot}~yr^{-1}$, $(g)$ median rest-frame $g-r$ colour, $(h)$ median $g$-band absolute magnitude. The final columns detail the dominant classified morphology (Hubble type or `St' for star or 'Sp' for spirals where specific Hubble type could not be decided) of each morphological cluster including any other notable features and the median silhouette score.}

\begin{tabular}{@{}llllllllllll@{}}
\toprule
\#  & $N^{a}$  & $N_{ext}\,^{b}$  & $z\,^{c}$  & $\langle  \mu_{g}  \rangle_{\mathrm{Kron}}\,^{d}$  & $log_{10}(M_{\star}/\mathrm{M}_{\odot})\,^{e}$  & $log_{10}(sSFR)^{f}$  & $(g\  -  \  r)^{g}$  & M$_{g}\,^{h}$  & comment  & score  \\
\midrule
0  & 1379  & 1122  & $  0.41  \pm    0.14$  & $23.05  \pm    0.41$  & $10.52  \pm    0.40$  & $-13.11  \pm    2.00$  & $  0.67  \pm    0.09$  & $-21.09  \pm    0.94$  & E  &   0.41   &  \\
1  & 110  & 110  & $  0.76  \pm    0.18$  & $24.60  \pm    0.55$  & $10.77  \pm    0.56$  & $-10.41  \pm    1.03$  & $  0.64  \pm    0.15$  & $-22.08  \pm    1.14$  & S0/Sa  &   0.11   &  \\
2  & 418  & 418  & $  0.72  \pm    0.24$  & $24.08  \pm    0.37$  & $10.02  \pm    0.46$  & $-9.45  \pm    0.55$  & $  0.37  \pm    0.10$  & $-21.36  \pm    1.13$  & S0/Sa  &   0.23   &  \\
3  & 267  & 267  & $  0.52  \pm    0.19$  & $23.70  \pm    0.33$  & $  9.94  \pm    0.47$  & $-9.53  \pm    0.64$  & $  0.39  \pm    0.11$  & $-20.74  \pm    2.89$  & Sp  &   0.13   &  \\
4  & 107  & 88  & $  0.31  \pm    0.16$  & $21.87  \pm    0.36$  & $10.17  \pm    0.41$  & $-9.62  \pm    1.28$  & $  0.45  \pm    0.12$  & $-20.81  \pm    4.58$  & E  &   0.07   &  \\
5  & 169  & 168  & $  0.28  \pm    0.33$  & $23.70  \pm    0.50$  & $  9.09  \pm    0.48$  & $-9.37  \pm    0.43$  & $  0.25  \pm    0.07$  & $-19.08  \pm    1.45$  & Sb/Sc  &   0.04   &  \\
6  & 519  & 507  & $  0.34  \pm    0.11$  & $22.55  \pm    0.29$  & $  9.89  \pm    0.43$  & $-9.53  \pm    0.47$  & $  0.39  \pm    0.10$  & $-20.54  \pm    1.15$  & Sa/Sd  &   0.24   &  \\
7  & 171  & 148  & $  0.35  \pm    0.06$  & $22.44  \pm    0.25$  & $10.73  \pm    0.21$  & $-13.65  \pm    1.77$  & $  0.70  \pm    0.05$  & $-21.35  \pm    0.56$  & S0  &   0.16   &  \\
8  & 87  & 83  & $  1.06  \pm    0.57$  & $25.80  \pm    1.40$  & $10.17  \pm    0.89$  & $-9.68  \pm    1.57$  & $  0.40  \pm    0.19$  & $-21.61  \pm    1.98$  & Sp,  diffuse  & -0.01   &  \\
9  & 443  & 440  & $  0.72  \pm    0.22$  & $24.29  \pm    0.54$  & $10.66  \pm    0.46$  & $-9.85  \pm    0.46$  & $  0.54  \pm    0.10$  & $-22.06  \pm    1.06$  & Sb/Sc  &   0.08   &  \\
10  & 253  & 253  & $  0.20  \pm    0.19$  & $22.71  \pm    0.43$  & $  9.47  \pm    0.46$  & $-9.52  \pm    0.38$  & $  0.33  \pm    0.10$  & $-19.43  \pm    1.43$  & Sc/Sd  &   0.05   &  \\
11  & 1383  & 1349  & $  0.54  \pm    0.16$  & $23.88  \pm    0.44$  & $10.62  \pm    0.49$  & $-11.52  \pm    1.95$  & $  0.67  \pm    0.09$  & $-21.42  \pm    2.08$  & E  &   0.48   &  \\
12  & 0  & $-$  & $-  \pm  -$  & $-  \pm  -$  & $-  \pm  -$  & $-  \pm  -$  & $-  \pm  -$  & $-  \pm  -$  & St  & $-$   &  \\
13  & 755  & 371  & $  0.32  \pm    0.31$  & $22.24  \pm    0.55$  & $  9.36  \pm    0.72$  & $-9.35  \pm    2.30$  & $  0.27  \pm    0.13$  & $-19.74  \pm    6.73$  & St  &   0.50   &  \\
14  & 308  & 299  & $  1.09  \pm    0.78$  & $24.63  \pm    0.73$  & $  9.52  \pm    1.12$  & $-9.26  \pm    0.49$  & $  0.23  \pm    0.12$  & $-21.00  \pm    3.04$  & Sp,  disturbed  &   0.38   &  \\
15  & 726  & 720  & $  0.29  \pm    0.47$  & $23.92  \pm    0.53$  & $  8.93  \pm    0.83$  & $-9.30  \pm    0.26$  & $  0.23  \pm    0.08$  & $-18.71  \pm    2.39$  & Sa/Sb/Sc,  many  edge-on  &   0.21   &  \\
16  & 1921  & 1918  & $  0.76  \pm    0.40$  & $24.35  \pm    0.56$  & $  9.79  \pm    0.72$  & $-9.38  \pm    0.53$  & $  0.32  \pm    0.12$  & $-21.07  \pm    1.79$  & Sp  &   0.34   &  \\
17  & 307  & 59  & $  0.04  \pm    0.68$  & $19.10  \pm    0.82$  & $  9.22  \pm    0.89$  & $  0.00  \pm    4.05$  & $  0.36  \pm    0.14$  & $-16.92  \pm  20.93$  & St  &   0.64   &  \\
18  & 210  & 209  & $  0.44  \pm    0.14$  & $23.06  \pm    0.65$  & $10.43  \pm    0.41$  & $-9.69  \pm    0.88$  & $  0.53  \pm    0.09$  & $-21.25  \pm    1.06$  & S0/Sa  &   0.01   &  \\
19  & 2513  & 2396  & $  0.46  \pm    0.24$  & $23.15  \pm    0.42$  & $  9.68  \pm    0.60$  & $-9.42  \pm    0.72$  & $  0.31  \pm    0.12$  & $-20.38  \pm    2.17$  & Sa/Sb/Sc  &   0.39   &  \\
20  & 83  & 4  & $  0.26  \pm    0.13$  & $20.47  \pm    1.33$  & $  9.05  \pm    0.96$  & $-9.13  \pm    3.99$  & $  0.23  \pm    0.06$  & $-19.88  \pm    3.31$  & St  &   0.32   &  \\
21  & 19  & 19  & $  0.09  \pm    0.64$  & $22.73  \pm    1.20$  & $  8.72  \pm    1.12$  & $-9.10  \pm    0.67$  & $  0.19  \pm    0.14$  & $-18.43  \pm    3.71$  & St  &   0.12   &  \\
22  & 996  & 139  & $  0.23  \pm    0.32$  & $21.55  \pm    0.73$  & $  9.45  \pm    1.10$  & $-9.33  \pm    3.14$  & $  0.28  \pm    0.16$  & $-19.60  \pm  10.42$  & St  &   0.83   &  \\
23  & 1  & 0  & $$-$  \pm  $-$$  & $$-$  \pm  $-$$  & $$-$  \pm  $-$$  & $$-$  \pm  $-$$  & $$-$  \pm  $-$$  & $$-$  \pm  $-$$  & St  & $-$   &  \\
24  & 29  & 7  & $  0.00  \pm    0.27$  & $21.19  \pm    0.56$  & $10.29  \pm    1.66$  & $  0.00  \pm    4.46$  & $  1.21  \pm    0.40$  & $21.31  \pm  19.45$  & St  &   0.31   &  \\
25  & 628  & 109  & $  0.41  \pm    0.25$  & $22.71  \pm    0.81$  & $10.83  \pm    0.54$  & $-14.62  \pm    2.47$  & $  0.70  \pm    0.21$  & $-21.56  \pm    4.52$  & St  &   0.65   &  \\
26  & 114  & 114  & $  0.14  \pm    0.08$  & $20.83  \pm    0.54$  & $10.45  \pm    0.52$  & $-10.48  \pm    2.09$  & $  0.64  \pm    0.12$  & $-20.43  \pm    3.92$  & Sp  &   0.09   &  \\
27  & 596  & 596  & $  0.42  \pm    0.23$  & $23.63  \pm    0.42$  & $  9.84  \pm    0.53$  & $-9.45  \pm    0.36$  & $  0.35  \pm    0.09$  & $-20.60  \pm    1.37$  & Sa/Sb  &   0.16   &  \\
28  & 215  & 213  & $  0.65  \pm    0.26$  & $23.93  \pm    0.32$  & $10.06  \pm    0.47$  & $-9.47  \pm    0.37$  & $  0.38  \pm    0.11$  & $-21.19  \pm    1.11$  & Sp,  asymmetries  &   0.06   &  \\
29  & 582  & 578  & $  0.91  \pm    0.66$  & $25.15  \pm    0.92$  & $  9.63  \pm    0.89$  & $-9.41  \pm    0.81$  & $  0.28  \pm    0.14$  & $-20.79  \pm    2.26$  & Sp,  diffuse  &   0.34   &  \\
30  & 276  & 275  & $  0.80  \pm    0.14$  & $24.10  \pm    0.41$  & $11.08  \pm    0.31$  & $-10.11  \pm    0.75$  & $  0.61  \pm    0.10$  & $-22.83  \pm    0.66$  & S0/Sa  &   0.27   &  \\
31  & 172  & 172  & $  0.29  \pm    0.07$  & $22.14  \pm    0.52$  & $10.96  \pm    0.25$  & $-14.65  \pm    1.80$  & $  0.73  \pm    0.05$  & $-21.53  \pm    0.68$  & E  &   0.10   &  \\
32  & 22  & 22  & $  0.50  \pm    0.26$  & $24.59  \pm    0.62$  & $  9.33  \pm    0.66$  & $-9.43  \pm    0.17$  & $  0.31  \pm    0.09$  & $-19.78  \pm    1.67$  & Sb/Sc,  asymmetries  &   0.32   &  \\
33  & 322  & 313  & $  0.49  \pm    0.13$  & $23.32  \pm    0.47$  & $10.50  \pm    0.47$  & $-10.48  \pm    2.21$  & $  0.62  \pm    0.09$  & $-21.32  \pm    1.09$  & S0/Sa  &   0.40   &  \\
34  & 112  & 110  & $  0.79  \pm    0.21$  & $25.24  \pm    0.75$  & $11.09  \pm    0.55$  & $-11.21  \pm    1.71$  & $  0.70  \pm    0.14$  & $-22.48  \pm    1.02$  & S0/Sa,  asymmetries  &   0.17   &  \\
35  & 229  & 229  & $  0.38  \pm    0.41$  & $23.53  \pm    0.52$  & $  9.17  \pm    0.89$  & $-9.24  \pm    0.33$  & $  0.22  \pm    0.10$  & $-19.48  \pm    2.52$  & Sb/Sc  &   0.19   &  \\
36  & 28  & 19  & $  0.08  \pm    0.59$  & $18.69  \pm    0.24$  & $  8.96  \pm    0.66$  & $  0.00  \pm    5.06$  & $  0.36  \pm    0.12$  & $-18.69  \pm  17.99$  & St  &   0.44   &  \\
37  & 22  & 16  & $  0.18  \pm    0.14$  & $19.82  \pm    0.78$  & $10.86  \pm    0.46$  & $-10.35  \pm    4.83$  & $  0.71  \pm    0.24$  & $-20.41  \pm  19.18$  & St  &   0.19   &  \\
38  & 97  & 94  & $  0.90  \pm    0.42$  & $25.19  \pm    0.79$  & $10.77  \pm    0.50$  & $-10.35  \pm    0.97$  & $  0.59  \pm    0.13$  & $-22.35  \pm    1.20$  & Sp,  red  &   0.06   &  \\
39  & 546  & 45  & $  0.21  \pm    0.37$  & $20.83  \pm    1.78$  & $  9.64  \pm    0.98$  & $-9.55  \pm    4.01$  & $  0.37  \pm    0.19$  & $-19.95  \pm  13.89$  & St  &   0.40   &  \\
40  & 178  & 152  & $  0.35  \pm    0.67$  & $23.02  \pm    0.74$  & $  8.81  \pm    1.28$  & $-9.08  \pm    0.96$  & $  0.15  \pm    0.13$  & $-19.15  \pm    3.91$  & E,  blue  &   0.60   &  \\
41  & 0  & $-$  & $-  \pm  -$  & $-  \pm  -$  & $-  \pm  -$  & $-  \pm  -$  & $-  \pm  -$  & $-  \pm  -$  & St  & $-$   &  \\
42  & 186  & 186  & $  0.36  \pm    0.10$  & $22.83  \pm    0.21$  & $10.74  \pm    0.37$  & $-10.46  \pm    1.64$  & $  0.67  \pm    0.06$  & $-21.35  \pm    0.95$  & Sa/Sb  &   0.12   &  \\
43  & 404  & 402  & $  0.56  \pm    0.15$  & $23.72  \pm    0.31$  & $10.81  \pm    0.44$  & $-10.03  \pm    0.96$  & $  0.62  \pm    0.09$  & $-21.97  \pm    1.04$  & Sa  &   0.14   &  \\
44  & 274  & 138  & $  0.00  \pm    0.40$  & $20.79  \pm    1.16$  & $11.77  \pm    0.82$  & $  0.00  \pm    5.45$  & $  1.21  \pm    0.36$  & $20.71  \pm  21.99$  & St  &   0.57   &  \\
45  & 275  & 270  & $  0.89  \pm    0.49$  & $25.02  \pm    0.67$  & $10.11  \pm    0.75$  & $-9.52  \pm    0.76$  & $  0.40  \pm    0.14$  & $-21.33  \pm    1.76$  & Sa/Sb/Sc  &   0.33   &  \\
46  & 275  & 20  & $  0.12  \pm    0.55$  & $20.76  \pm    0.88$  & $  8.88  \pm    1.14$  & $-9.08  \pm    4.09$  & $  0.25  \pm    0.21$  & $-18.11  \pm  12.01$  & St  &   0.63   &  \\
47  & 544  & 522  & $  0.74  \pm    0.35$  & $24.81  \pm    0.59$  & $10.91  \pm    0.52$  & $-11.01  \pm    1.67$  & $  0.68  \pm    0.15$  & $-22.23  \pm    2.39$  & E  &   0.46   &  \\
48  & 179  & 179  & $  0.20  \pm    0.13$  & $22.83  \pm    0.29$  & $  8.90  \pm    0.54$  & $-9.37  \pm    0.24$  & $  0.23  \pm    0.07$  & $-18.47  \pm    1.60$  & Sc/Sd  &   0.16   &  \\
49  & 192  & 192  & $  0.84  \pm    0.30$  & $24.64  \pm    0.78$  & $10.64  \pm    0.46$  & $-9.84  \pm    1.24$  & $  0.51  \pm    0.13$  & $-22.13  \pm    1.00$  & Sa/Sb  &   0.44   &  \\
50  & 191  & 191  & $  0.18  \pm    0.18$  & $21.80  \pm    0.30$  & $  9.27  \pm    0.53$  & $-9.39  \pm    0.79$  & $  0.26  \pm    0.09$  & $-19.33  \pm    1.76$  & Sp  &   0.16   &  \\
51  & 248  & 248  & $  0.36  \pm    0.11$  & $22.63  \pm    0.27$  & $10.49  \pm    0.34$  & $-9.80  \pm    1.08$  & $  0.56  \pm    0.09$  & $-21.11  \pm    1.08$  & Sa  &   0.13   &  \\
52  & 271  & 270  & $  0.42  \pm    0.29$  & $23.85  \pm    0.39$  & $  9.41  \pm    0.70$  & $-9.38  \pm    0.25$  & $  0.28  \pm    0.10$  & $-19.89  \pm    1.83$  & Sb/Sc  &   0.32   &  \\
53  & 363  & 362  & $  0.28  \pm    0.22$  & $22.95  \pm    0.44$  & $  9.29  \pm    0.74$  & $-9.40  \pm    0.23$  & $  0.27  \pm    0.11$  & $-19.51  \pm    1.88$  & Sa/Sb  &   0.28   &  \\
54  & 456  & 454  & $  0.47  \pm    0.14$  & $23.49  \pm    0.34$  & $10.83  \pm    0.45$  & $-10.35  \pm    1.30$  & $  0.67  \pm    0.09$  & $-21.73  \pm    1.09$  & S0/Sa  &   0.05   &  \\
55  & 1  & 1  & $  0.05  \pm    0.00$  & $24.13  \pm    0.00$  & $  7.47  \pm    0.00$  & $-9.42  \pm    0.00$  & $  0.25  \pm    0.00$  & $-14.34  \pm    0.00$  & St  &   0.04   &  \\
56  & 4  & 4  & $  0.30  \pm    0.29$  & $20.14  \pm    0.38$  & $11.26  \pm    1.48$  & $-10.06  \pm    4.43$  & $  0.55  \pm    0.22$  & $-19.59  \pm  17.41$  & St  &   0.38   &  \\
57  & 260  & 256  & $  1.01  \pm    0.41$  & $24.82  \pm    0.72$  & $10.84  \pm    0.61$  & $-10.00  \pm    0.92$  & $  0.56  \pm    0.14$  & $-22.69  \pm    1.38$  & S0/Sa  &   0.41   &  \\
58  & 173  & 173  & $  0.49  \pm    0.38$  & $24.12  \pm    0.48$  & $  9.50  \pm    0.79$  & $-9.39  \pm    0.29$  & $  0.29  \pm    0.09$  & $-20.28  \pm    2.15$  & Sb/Sc  &   0.20   &  \\
\bottomrule
\end{tabular}

\label{table:props}
\end{table*}

\begin{table*}
\contcaption{}
\begin{tabular}{@{}llllllllllll@{}}
\toprule
\#  & $N^{a}$  & $N_{ext}\,^{b}$  & $z\,^{c}$  & $\langle  \mu_{g}  \rangle_{\mathrm{Kron}}\,^{d}$  & $log_{10}(M_{\star}/\mathrm{M}_{\odot})\,^{e}$  & $log_{10}(sSFR)^{f}$  & $(g\  -  \  r)^{g}$  & M$_{g}\,^{h}$  & comment  & score  \\
\midrule
59  & 40  & 40  & $  0.12  \pm    0.24$  & $23.09  \pm    1.41$  & $  9.44  \pm    0.85$  & $-9.63  \pm    1.25$  & $  0.39  \pm    0.15$  & $-19.02  \pm    2.16$  & St  & -0.01   &  \\
60  & 85  & 80  & $  0.87  \pm    0.67$  & $25.32  \pm    0.87$  & $  9.60  \pm    0.91$  & $-9.44  \pm    0.51$  & $  0.30  \pm    0.15$  & $-20.60  \pm    2.27$  & Sp,  diffuse  &   0.38   &  \\
61  & 2184  & 2176  & $  0.66  \pm    0.22$  & $23.98  \pm    0.39$  & $10.17  \pm    0.52$  & $-9.62  \pm    0.70$  & $  0.45  \pm    0.12$  & $-21.30  \pm    1.56$  & E  &   0.41   &  \\
62  & 111  & 109  & $  0.25  \pm    0.21$  & $21.92  \pm    1.00$  & $10.96  \pm    0.38$  & $-14.76  \pm    1.83$  & $  0.73  \pm    0.09$  & $-21.54  \pm    0.83$  & St  &   0.02   &  \\
63  & 3  & 3  & $  1.49  \pm    0.54$  & $24.39  \pm    1.00$  & $11.59  \pm    0.57$  & $-11.31  \pm    0.89$  & $  0.28  \pm    0.06$  & $-26.10  \pm    2.11$  & Sb/Sc  &   0.39   &  \\
64  & 196  & 196  & $  0.65  \pm    0.21$  & $24.31  \pm    0.44$  & $  9.89  \pm    0.42$  & $-9.41  \pm    0.32$  & $  0.34  \pm    0.09$  & $-21.02  \pm    1.02$  & St  &   0.08   &  \\
65  & 91  & 43  & $  0.00  \pm    0.34$  & $21.30  \pm    0.20$  & $11.98  \pm    0.03$  & $  0.00  \pm    4.55$  & $  1.37  \pm    0.30$  & $21.58  \pm  18.31$  & Sa  &   0.74   &  \\
66  & 162  & 160  & $  0.21  \pm    0.09$  & $22.02  \pm    0.44$  & $10.68  \pm    0.35$  & $-13.13  \pm    2.05$  & $  0.73  \pm    0.08$  & $-20.60  \pm    0.94$  & St  &   0.06   &  \\
67  & 1  & 1  & $  0.02  \pm    0.00$  & $    nan  \pm      nan$  & $  7.45  \pm    0.00$  & $-11.45  \pm    0.00$  & $  0.54  \pm    0.00$  & $-12.37  \pm    0.00$  & S0/Sa  & -0.01   &  \\
68  & 102  & 100  & $  0.89  \pm    0.36$  & $25.18  \pm    0.95$  & $10.90  \pm    0.53$  & $-10.59  \pm    1.43$  & $  0.64  \pm    0.13$  & $-22.49  \pm    1.20$  & Sa/Sb/Sc  &   0.39   &  \\
69  & 420  & 419  & $  0.36  \pm    0.15$  & $22.62  \pm    0.40$  & $  9.81  \pm    0.42$  & $-9.48  \pm    0.42$  & $  0.35  \pm    0.10$  & $-20.43  \pm    1.19$  & Sc/Sd  &   0.05   &  \\
70  & 202  & 199  & $  0.62  \pm    0.51$  & $23.55  \pm    0.58$  & $  9.28  \pm    1.07$  & $-9.17  \pm    0.52$  & $  0.21  \pm    0.11$  & $-20.74  \pm    3.06$  & Sp,  asymmetries  &   0.14   &  \\
71  & 358  & 334  & $  0.22  \pm    0.11$  & $21.72  \pm    0.39$  & $10.27  \pm    0.42$  & $-9.83  \pm    1.21$  & $  0.53  \pm    0.12$  & $-20.39  \pm    1.32$  & S0/Sa  &   0.18   &  \\
72  & 57  & 56  & $  0.12  \pm    0.12$  & $20.76  \pm    1.97$  & $10.57  \pm    0.67$  & $-13.64  \pm    2.19$  & $  0.74  \pm    0.16$  & $-20.42  \pm    1.23$  & E  &   0.11   &  \\
73  & 0  & $-$  & $-  \pm  -$  & $-  \pm  -$  & $-  \pm  -$  & $-  \pm  -$  & $-  \pm  -$  & $-  \pm  -$  & St  & $-$   &  \\
74  & 507  & 507  & $  0.70  \pm    0.29$  & $24.47  \pm    0.42$  & $  9.78  \pm    0.64$  & $-9.39  \pm    0.43$  & $  0.32  \pm    0.11$  & $-20.94  \pm    1.53$  & Sb/Sc  &   0.11   &  \\
75  & 128  & 125  & $  0.82  \pm    0.31$  & $23.42  \pm    0.69$  & $10.02  \pm    0.77$  & $-9.23  \pm    0.61$  & $  0.26  \pm    0.11$  & $-22.02  \pm    2.22$  & Sa/Sb  &   0.45   &  \\
76  & 71  & 71  & $  0.30  \pm    0.79$  & $24.37  \pm    0.70$  & $  8.72  \pm    1.09$  & $-9.22  \pm    0.28$  & $  0.21  \pm    0.11$  & $-18.27  \pm    3.09$  & Sc/Sd  &   0.45   &  \\
77  & 0  & $-$  & $-  \pm  -$  & $-  \pm  -$  & $-  \pm  -$  & $-  \pm  -$  & $-  \pm  -$  & $-  \pm  -$  & St  & $-$   &  \\
78  & 194  & 194  & $  0.16  \pm    0.06$  & $21.55  \pm    0.33$  & $  9.79  \pm    0.38$  & $-9.64  \pm    0.81$  & $  0.40  \pm    0.12$  & $-19.86  \pm    0.79$  & Sa/Sb  &   0.09   &  \\
79  & 314  & 314  & $  0.25  \pm    0.41$  & $23.66  \pm    0.39$  & $  8.75  \pm    0.73$  & $-9.29  \pm    0.28$  & $  0.21  \pm    0.07$  & $-18.27  \pm    2.21$  & Sc/Sd  &   0.22   &  \\
80  & 147  & 147  & $  1.03  \pm    0.23$  & $25.16  \pm    0.72$  & $10.64  \pm    0.53$  & $-10.00  \pm    0.97$  & $  0.51  \pm    0.13$  & $-22.26  \pm    1.13$  & Sp,  red  &   0.37   &  \\
81  & 419  & 417  & $  0.52  \pm    0.27$  & $23.49  \pm    0.37$  & $10.38  \pm    0.43$  & $-9.67  \pm    0.51$  & $  0.50  \pm    0.09$  & $-21.29  \pm    1.11$  & S0/Sa  &   0.10   &  \\
82  & 843  & 842  & $  0.68  \pm    0.20$  & $24.26  \pm    0.42$  & $10.49  \pm    0.43$  & $-9.92  \pm    0.57$  & $  0.54  \pm    0.10$  & $-21.67  \pm    1.06$  & S0/Sa  &   0.19   &  \\
83  & 15  & 13  & $  0.70  \pm    0.46$  & $24.52  \pm    1.40$  & $11.20  \pm    0.44$  & $-11.81  \pm    2.34$  & $  0.68  \pm    0.18$  & $-22.49  \pm    0.63$  & St  & -0.03   &  \\
84  & 229  & 229  & $  0.38  \pm    0.17$  & $22.89  \pm    0.29$  & $10.09  \pm    0.34$  & $-9.55  \pm    0.58$  & $  0.42  \pm    0.10$  & $-20.92  \pm    0.94$  & Sa  &   0.19   &  \\
85  & 0  & $-$  & $-  \pm  -$  & $-  \pm  -$  & $-  \pm  -$  & $-  \pm  -$  & $-  \pm  -$  & $-  \pm  -$  & St  & $-$   &  \\
86  & 276  & 267  & $  0.22  \pm    0.16$  & $22.62  \pm    0.46$  & $  9.13  \pm    0.47$  & $-9.39  \pm    0.61$  & $  0.25  \pm    0.08$  & $-19.05  \pm    2.87$  & S0/Sa,  blue  &   0.16   &  \\
87  & 35  & 33  & $  0.31  \pm    0.84$  & $24.47  \pm    1.49$  & $  9.41  \pm    0.90$  & $-9.39  \pm    0.62$  & $  0.26  \pm    0.16$  & $-20.04  \pm    2.82$  & E,  blue  &   0.15   &  \\
88  & 131  & 131  & $  0.13  \pm    0.06$  & $21.09  \pm    0.76$  & $10.61  \pm    0.44$  & $-11.45  \pm    2.01$  & $  0.71  \pm    0.10$  & $-20.36  \pm    1.15$  & S0/Sa  &   0.05   &  \\
89  & 2  & 2  & $  1.12  \pm    0.00$  & $    nan  \pm      nan$  & $11.33  \pm    0.00$  & $-11.31  \pm    0.00$  & $  0.31  \pm    0.00$  & $-25.24  \pm    0.00$  & St  &   0.35   &  \\
90  & 430  & 429  & $  0.34  \pm    0.27$  & $23.43  \pm    0.45$  & $  9.56  \pm    0.62$  & $-9.44  \pm    0.31$  & $  0.31  \pm    0.10$  & $-20.05  \pm    1.62$  & Sc/Sd  &   0.06   &  \\
91  & 436  & 434  & $  0.33  \pm    0.13$  & $22.65  \pm    0.39$  & $10.14  \pm    0.37$  & $-9.62  \pm    0.40$  & $  0.46  \pm    0.09$  & $-20.87  \pm    1.07$  & Sb/Sc  &   0.00   &  \\
92  & 3  & 3  & $  0.55  \pm    0.06$  & $21.78  \pm    0.56$  & $11.41  \pm    0.35$  & $-14.62  \pm    2.45$  & $  0.76  \pm    0.16$  & $-22.98  \pm    0.19$  & St  &   0.43   &  \\
93  & 93  & 91  & $  1.19  \pm    0.98$  & $25.28  \pm    1.10$  & $  9.15  \pm    1.06$  & $-9.26  \pm    0.52$  & $  0.23  \pm    0.14$  & $-20.05  \pm    2.94$  & Sb/Sc/Sd,  diffuse  &   0.35   &  \\
94  & 235  & 235  & $  0.18  \pm    0.08$  & $22.17  \pm    0.46$  & $10.26  \pm    0.38$  & $-9.95  \pm    1.70$  & $  0.56  \pm    0.12$  & $-20.16  \pm    0.99$  & Sb/Sc  &   0.10   &  \\
95  & 205  & 171  & $  0.34  \pm    0.12$  & $22.43  \pm    0.45$  & $10.34  \pm    0.36$  & $-10.16  \pm    1.69$  & $  0.62  \pm    0.10$  & $-20.92  \pm    1.04$  & E  &   0.22   &  \\
96  & 900  & 893  & $  0.44  \pm    0.11$  & $23.27  \pm    0.42$  & $10.74  \pm    0.39$  & $-10.75  \pm    1.85$  & $  0.68  \pm    0.09$  & $-21.46  \pm    0.92$  & S0/Sa  &   0.11   &  \\
97  & 152  & 152  & $  0.29  \pm    0.11$  & $22.56  \pm    0.29$  & $10.74  \pm    0.36$  & $-10.29  \pm    1.93$  & $  0.70  \pm    0.10$  & $-21.17  \pm    1.10$  & Sa  &   0.02   &  \\
98  & 284  & 284  & $  0.46  \pm    0.29$  & $23.79  \pm    0.46$  & $  9.56  \pm    0.66$  & $-9.37  \pm    0.28$  & $  0.29  \pm    0.10$  & $-20.29  \pm    1.73$  & E  &   0.48   &  \\
99  & 602  & 593  & $  0.54  \pm    0.24$  & $23.88  \pm    0.49$  & $10.72  \pm    0.43$  & $-10.71  \pm    1.69$  & $  0.67  \pm    0.11$  & $-21.63  \pm    0.98$  & Sp,  clumpy  &   0.16   &  \\
100  & 55  & 12  & $  0.71  \pm    0.43$  & $24.46  \pm    1.20$  & $10.35  \pm    1.04$  & $-9.70  \pm    1.71$  & $  0.42  \pm    0.22$  & $-21.86  \pm    2.09$  & St  &   0.69   &  \\
101  & 161  & 161  & $  0.18  \pm    0.06$  & $21.28  \pm    0.23$  & $10.44  \pm    0.33$  & $-11.79  \pm    1.96$  & $  0.68  \pm    0.09$  & $-20.42  \pm    0.85$  & Sa  &   0.13   &  \\
102  & 23  & 22  & $  0.07  \pm    0.39$  & $20.38  \pm    1.23$  & $10.49  \pm    0.56$  & $-11.64  \pm    2.23$  & $  0.74  \pm    0.17$  & $-19.67  \pm    1.42$  & saturated  &   0.16   &  \\
103  & 244  & 244  & $  0.51  \pm    0.31$  & $23.88  \pm    0.37$  & $  9.57  \pm    0.53$  & $-9.32  \pm    0.25$  & $  0.29  \pm    0.08$  & $-20.44  \pm    1.43$  & Sb/Sc/Sd  &   0.11   &  \\
104  & 350  & 347  & $  0.38  \pm    0.19$  & $23.15  \pm    0.39$  & $  9.79  \pm    0.65$  & $-9.42  \pm    0.76$  & $  0.33  \pm    0.09$  & $-20.50  \pm    1.72$  & Sb/Sc  &   0.12   &  \\
105  & 0  & $-$  & $-  \pm  -$  & $-  \pm  -$  & $-  \pm  -$  & $-  \pm  -$  & $-  \pm  -$  & $-  \pm  -$  & St  & $-$   &  \\
106  & 0  & $-$  & $-  \pm  -$  & $-  \pm  -$  & $-  \pm  -$  & $-  \pm  -$  & $-  \pm  -$  & $-  \pm  -$  & St  & $-$   &  \\
107  & 60  & 56  & $  1.10  \pm    0.66$  & $25.94  \pm    1.33$  & $10.52  \pm    0.62$  & $-10.47  \pm    1.85$  & $  0.55  \pm    0.18$  & $-22.01  \pm    1.75$  & E  &   0.46   &  \\
108  & 0  & $-$  & $-  \pm  -$  & $-  \pm  -$  & $-  \pm  -$  & $-  \pm  -$  & $-  \pm  -$  & $-  \pm  -$  & St  & $-$   &  \\
109  & 152  & 152  & $  0.15  \pm    0.07$  & $21.93  \pm    0.40$  & $  9.57  \pm    0.59$  & $-9.54  \pm    0.85$  & $  0.33  \pm    0.13$  & $-19.56  \pm    1.53$  & Sp,  clumpy  &   0.11   &  \\
110  & 73  & 71  & $  0.43  \pm    0.21$  & $23.16  \pm    0.65$  & $10.28  \pm    0.56$  & $-9.59  \pm    1.32$  & $  0.44  \pm    0.14$  & $-21.21  \pm    1.54$  & Sb/Sc  &   0.26   &  \\
111  & 4  & 3  & $  1.95  \pm    0.65$  & $25.74  \pm    0.84$  & $  9.92  \pm    0.26$  & $-9.00  \pm    0.09$  & $  0.13  \pm    0.05$  & $-22.63  \pm    0.62$  & S0/Sa  &   0.29   &  \\
112  & 80  & 78  & $  0.80  \pm    0.16$  & $24.19  \pm    0.51$  & $11.16  \pm    0.57$  & $-10.38  \pm    0.81$  & $  0.64  \pm    0.15$  & $-22.82  \pm    0.96$  & saturated  &   0.32   &  \\
113  & 719  & 49  & $  0.37  \pm    0.37$  & $21.52  \pm    1.27$  & $11.16  \pm    1.14$  & $-9.37  \pm    5.19$  & $  0.76  \pm    0.45$  & $-20.56  \pm  21.43$  & Sa/Sb,  red  &   0.73   &  \\
114  & 344  & 325  & $  0.34  \pm    0.19$  & $22.42  \pm    0.47$  & $  9.65  \pm    0.58$  & $-9.45  \pm    0.78$  & $  0.33  \pm    0.10$  & $-20.15  \pm    2.83$  & St  &   0.22   &  \\
115  & 45  & 34  & $  0.13  \pm    0.86$  & $19.02  \pm    0.31$  & $  9.61  \pm    0.59$  & $  0.00  \pm    4.48$  & $  0.41  \pm    0.20$  & $-19.86  \pm  20.14$  & saturated  &   0.27   &  \\
116  & 312  & 312  & $  0.33  \pm    0.15$  & $23.05  \pm    0.41$  & $  9.97  \pm    0.47$  & $-9.58  \pm    0.93$  & $  0.41  \pm    0.12$  & $-20.49  \pm    1.19$  & St  &   0.06   &  \\
117  & 171  & 171  & $  0.58  \pm    0.52$  & $24.46  \pm    0.60$  & $  9.22  \pm    0.92$  & $-9.27  \pm    0.57$  & $  0.23  \pm    0.12$  & $-19.93  \pm    2.41$  & Sb/Sc  &   0.09   &  \\
118  & 262  & 261  & $  0.48  \pm    0.14$  & $23.28  \pm    0.25$  & $10.47  \pm    0.35$  & $-9.66  \pm    0.33$  & $  0.50  \pm    0.07$  & $-21.50  \pm    0.94$  & Sb/Sc/Sd  &   0.15   &  \\
119  & 442  & 441  & $  0.48  \pm    0.38$  & $24.10  \pm    0.36$  & $  9.31  \pm    0.75$  & $-9.30  \pm    0.33$  & $  0.25  \pm    0.09$  & $-19.95  \pm    2.08$  & S0/Sa,  asymmetries,Sb/Sc  &   0.13   &  \\
120  & 155  & 77  & $  0.24  \pm    0.10$  & $21.55  \pm    0.35$  & $10.70  \pm    0.23$  & $-14.28  \pm    4.42$  & $  0.71  \pm    0.19$  & $-20.78  \pm  12.86$  & E  &   0.34   &  \\
121  & 141  & 141  & $  0.82  \pm    0.42$  & $24.49  \pm    0.43$  & $  9.72  \pm    0.80$  & $-9.37  \pm    0.27$  & $  0.28  \pm    0.10$  & $-21.08  \pm    2.07$  & Sp,  companions  & -0.01   &  \\
122  & 123  & 123  & $  0.11  \pm    0.05$  & $21.04  \pm    0.44$  & $  9.78  \pm    0.65$  & $-9.61  \pm    1.13$  & $  0.38  \pm    0.14$  & $-19.87  \pm    1.50$  & Sa/Sb/Sd,  blue  ring  &   0.09   &  \\
\bottomrule
\end{tabular}

\end{table*}

\begin{table*}
\contcaption{}
\begin{tabular}{@{}llllllllllll@{}}
\toprule
\#  & $N^{a}$  & $N_{ext}\,^{b}$  & $z\,^{c}$  & $\langle  \mu_{g}  \rangle_{\mathrm{Kron}}\,^{d}$  & $log_{10}(M_{\star}/\mathrm{M}_{\odot})\,^{e}$  & $log_{10}(sSFR)^{f}$  & $(g\  -  \  r)^{g}$  & M$_{g}\,^{h}$  & comment  & score  \\
\midrule
123  & 0  & $-$  & $-  \pm  -$  & $-  \pm  -$  & $-  \pm  -$  & $-  \pm  -$  & $-  \pm  -$  & $-  \pm  -$  & St  & $-$   &  \\
124  & 12  & 7  & $  0.00  \pm    0.29$  & $20.20  \pm    0.22$  & $11.62  \pm    0.23$  & $  0.00  \pm    4.77$  & $  1.20  \pm    0.20$  & $20.26  \pm  21.78$  & St  &   0.58   &  \\
125  & 1  & 1  & $  0.28  \pm    0.00$  & $19.87  \pm    0.00$  & $11.34  \pm    0.00$  & $-10.72  \pm    0.00$  & $  0.80  \pm    0.00$  & $-22.08  \pm    0.00$  & St  &   0.18   &  \\
126  & 209  & 208  & $  0.31  \pm    0.09$  & $22.36  \pm    0.24$  & $10.50  \pm    0.32$  & $-10.59  \pm    1.72$  & $  0.65  \pm    0.08$  & $-21.02  \pm    0.93$  & E  &   0.12   &  \\
127  & 158  & 154  & $  0.13  \pm    0.06$  & $22.59  \pm    0.41$  & $  9.05  \pm    0.69$  & $-9.38  \pm    0.38$  & $  0.25  \pm    0.11$  & $-18.94  \pm    1.76$  & Sc/Sd  &   0.18   &  \\
128  & 146  & 146  & $  0.11  \pm    0.29$  & $23.20  \pm    0.89$  & $  8.57  \pm    0.85$  & $-9.32  \pm    0.47$  & $  0.21  \pm    0.11$  & $-17.83  \pm    2.48$  & Sd  &   0.11   &  \\
129  & 301  & 275  & $  0.47  \pm    0.62$  & $23.99  \pm    1.80$  & $10.02  \pm    1.16$  & $-9.48  \pm    2.43$  & $  0.35  \pm    0.24$  & $-20.86  \pm    7.33$  & Sp  & -0.08   &  \\
130  & 193  & 193  & $  0.35  \pm    0.08$  & $22.73  \pm    0.34$  & $10.99  \pm    0.43$  & $-12.47  \pm    2.01$  & $  0.73  \pm    0.07$  & $-21.75  \pm    1.04$  & E  &   0.05   &  \\
131  & 1  & 1  & $  1.53  \pm    0.00$  & $24.35  \pm    0.00$  & $11.38  \pm    0.00$  & $-10.04  \pm    0.00$  & $  0.63  \pm    0.00$  & $-24.35  \pm    0.00$  & St  &   0.01   &  \\
132  & 183  & 174  & $  0.26  \pm    0.07$  & $22.04  \pm    0.30$  & $10.62  \pm    0.33$  & $-14.23  \pm    1.74$  & $  0.70  \pm    0.06$  & $-20.76  \pm    0.84$  & E  &   0.10   &  \\
133  & 1  & 0  & $$-$  \pm  $-$$  & $$-$  \pm  $-$$  & $$-$  \pm  $-$$  & $$-$  \pm  $-$$  & $$-$  \pm  $-$$  & $$-$  \pm  $-$$  & saturated  & $-$   &  \\
134  & 271  & 238  & $  0.18  \pm    0.25$  & $22.72  \pm    0.42$  & $  8.94  \pm    0.70$  & $-9.36  \pm    1.12$  & $  0.23  \pm    0.10$  & $-18.60  \pm    4.32$  & Sp  &   0.18   &  \\
135  & 82  & 81  & $  1.24  \pm    0.32$  & $25.20  \pm    1.19$  & $10.62  \pm    0.60$  & $-9.47  \pm    1.32$  & $  0.45  \pm    0.16$  & $-22.64  \pm    1.26$  & Sp  &   0.23   &  \\
136  & 1  & 1  & $  1.07  \pm    0.00$  & $27.54  \pm    0.00$  & $  9.53  \pm    0.00$  & $-9.61  \pm    0.00$  & $  0.31  \pm    0.00$  & $-20.33  \pm    0.00$  & saturated  &   0.14   &  \\
137  & 1  & 1  & $  0.02  \pm    0.00$  & $20.93  \pm    0.00$  & $  7.45  \pm    0.00$  & $-11.45  \pm    0.00$  & $  0.54  \pm    0.00$  & $-12.39  \pm    0.00$  & St  &   0.16   &  \\
138  & 931  & 930  & $  0.40  \pm    0.20$  & $23.52  \pm    0.33$  & $  9.67  \pm    0.53$  & $-9.44  \pm    0.27$  & $  0.33  \pm    0.09$  & $-20.19  \pm    1.36$  & Sa/Sb/Sc  &   0.20   &  \\
139  & 141  & 141  & $  0.63  \pm    0.20$  & $24.47  \pm    0.47$  & $10.85  \pm    0.37$  & $-11.21  \pm    1.57$  & $  0.69  \pm    0.11$  & $-21.98  \pm    0.80$  & E,  asymmetries  &   0.15   &  \\
140  & 148  & 141  & $  0.42  \pm    0.08$  & $23.06  \pm    0.35$  & $10.92  \pm    0.38$  & $-14.58  \pm    1.87$  & $  0.71  \pm    0.08$  & $-21.70  \pm    0.72$  & E,  companions  &   0.13   &  \\
141  & 154  & 154  & $  0.33  \pm    0.29$  & $23.96  \pm    0.38$  & $  8.99  \pm    0.77$  & $-9.35  \pm    0.47$  & $  0.25  \pm    0.08$  & $-18.85  \pm    2.18$  & Sp,  companions  &   0.22   &  \\
142  & 2  & 2  & $  0.21  \pm    0.17$  & $27.62  \pm    0.00$  & $  8.11  \pm    0.15$  & $-10.46  \pm    1.49$  & $  0.30  \pm    0.23$  & $-16.26  \pm    1.71$  & saturated  &   0.04   &  \\
143  & 117  & 116  & $  0.91  \pm    0.58$  & $24.76  \pm    0.76$  & $10.05  \pm    0.95$  & $-9.35  \pm    0.81$  & $  0.31  \pm    0.13$  & $-21.46  \pm    2.38$  & Sb/Sd/Sc  &   0.14   &  \\
144  & 129  & 129  & $  0.95  \pm    0.23$  & $24.18  \pm    0.46$  & $10.63  \pm    0.48$  & $-9.37  \pm    0.55$  & $  0.44  \pm    0.12$  & $-22.67  \pm    1.01$  & Sp,  clumpy,  red  &   0.12   &  \\
145  & 59  & 58  & $  0.89  \pm    0.34$  & $25.28  \pm    0.55$  & $10.45  \pm    0.69$  & $-9.67  \pm    1.21$  & $  0.49  \pm    0.15$  & $-21.62  \pm    1.69$  & S0  &   0.04   &  \\
146  & 258  & 257  & $  0.74  \pm    0.32$  & $24.57  \pm    0.60$  & $  9.86  \pm    0.68$  & $-9.39  \pm    0.27$  & $  0.32  \pm    0.11$  & $-21.15  \pm    1.59$  & Sa/Sb/Sc,  companions  &   0.09   &  \\
147  & 400  & 64  & $  0.31  \pm    0.26$  & $22.00  \pm    0.75$  & $10.42  \pm    0.41$  & $-12.13  \pm    1.84$  & $  0.65  \pm    0.13$  & $-20.87  \pm    1.29$  & St  &   0.56   &  \\
148  & 248  & 64  & $  0.00  \pm    0.83$  & $20.04  \pm    0.69$  & $10.38  \pm    0.33$  & $  0.00  \pm    5.62$  & $  0.74  \pm    0.27$  & $19.82  \pm  21.41$  & St  &   0.38   &  \\
149  & 0  & $-$  & $-  \pm  -$  & $-  \pm  -$  & $-  \pm  -$  & $-  \pm  -$  & $-  \pm  -$  & $-  \pm  -$  & St  & $-$   &  \\
150  & 1  & 1  & $  0.15  \pm    0.00$  & $20.58  \pm    0.00$  & $  8.67  \pm    0.00$  & $-8.98  \pm    0.00$  & $  0.09  \pm    0.00$  & $-18.63  \pm    0.00$  & St  &   0.10   &  \\
151  & 252  & 252  & $  0.39  \pm    0.09$  & $22.88  \pm    0.41$  & $11.03  \pm    0.26$  & $-14.65  \pm    1.81$  & $  0.73  \pm    0.05$  & $-21.90  \pm    0.61$  & E  &   0.09   &  \\
152  & 1307  & 1303  & $  0.42  \pm    0.20$  & $23.36  \pm    0.46$  & $  9.64  \pm    0.50$  & $-9.42  \pm    0.28$  & $  0.31  \pm    0.10$  & $-20.20  \pm    1.26$  & Sp,  companions  &   0.24   &  \\
153  & 88  & 86  & $  1.11  \pm    0.83$  & $24.86  \pm    1.06$  & $  9.78  \pm    1.08$  & $-9.34  \pm    0.45$  & $  0.25  \pm    0.11$  & $-21.28  \pm    3.02$  & Sd,  asymmetries  &   0.01   &  \\
154  & 39  & 18  & $  0.70  \pm    0.33$  & $24.44  \pm    0.57$  & $11.42  \pm    1.29$  & $-11.63  \pm    2.44$  & $  0.71  \pm    0.19$  & $-23.05  \pm    2.73$  & St  &   0.64   &  \\
155  & 244  & 244  & $  0.22  \pm    0.15$  & $22.11  \pm    0.29$  & $  9.64  \pm    0.45$  & $-9.51  \pm    0.28$  & $  0.35  \pm    0.10$  & $-19.67  \pm    1.23$  & Sa/Sb  &   0.14   &  \\
156  & 116  & 107  & $  0.60  \pm    0.13$  & $23.91  \pm    0.38$  & $11.17  \pm    0.50$  & $-14.91  \pm    1.87$  & $  0.70  \pm    0.09$  & $-22.52  \pm    1.08$  & E  &   0.10   &  \\
157  & 141  & 140  & $  0.88  \pm    0.47$  & $25.28  \pm    0.79$  & $  9.76  \pm    0.69$  & $-9.44  \pm    0.77$  & $  0.30  \pm    0.13$  & $-20.93  \pm    1.77$  & Sb/Sc/Sd,  asymmetries  &   0.14   &  \\
158  & 0  & $-$  & $-  \pm  -$  & $-  \pm  -$  & $-  \pm  -$  & $-  \pm  -$  & $-  \pm  -$  & $-  \pm  -$  & St  & $-$   &  \\
159  & 353  & 353  & $  0.82  \pm    0.37$  & $24.71  \pm    0.59$  & $  9.97  \pm    0.63$  & $-9.46  \pm    0.70$  & $  0.36  \pm    0.12$  & $-21.23  \pm    1.53$  & Sp  & -0.01   &  \\
\bottomrule
\end{tabular}

\label{tab:cluster_properties}
\end{table*}

\begin{table*}
 \caption{Example of 10 entries from the catalogue showing the position and morphological cluster membership individual galaxies. Columns are as follows: $(a)$ the RA of the centroid from the UML detection, $(b)$ the declination of the centroid from the UML detection, $(c)$ the RA of the centroid for the matched HSC object, $(d)$ the declination of the centroid for the matched HSC object, $(e)$ the ID of the matched HSC object, $(f)$ whether the matched HSC object is extended or not, $(g)$ the morphological cluster membership of the object, $(h)$ the number of pixels that make up the UML detection, $(i)$ the silhouette score for the object. Where there is no matching object in within $0.8''$ in the HSC catalogue, the 3rd to 6th columns are left blank.}
 \begin{tabular}{@{}lllllllll@{}}
\toprule
UML$_{\rm  RA}\,^{a}$  & UML$_{\rm  DEC}\,^{b}$  & HSC$_{\rm  RA}\,^{c}$  & HSC$_{\rm  DEC}\,^{d}$  & HSC  ID$\,^{e}$  & extended$\,^{f}$  & cluster  $\#\,^{g}$  & $n_{pix}\,^{h}$  & score$\,^{i}$  \\
\midrule
35.0484  & -5.3316  & $-$  & $-$  & $-$  & $-$  & 33  & 116  &   0.16  \\
35.0484  & -5.4808  & 35.0485  & -5.4808  & 37489923418246937  & False  & 132  & 112  & -0.10  \\
35.0480  & -5.4613  & 35.0479  & -5.4612  & 37489923418227875  & True  & 98  & 19  &   0.11  \\
35.0472  & -5.3657  & 35.0471  & -5.3657  & 37489923418254158  & True  & 91  & 232  &   0.08  \\
35.0475  & -5.3088  & 35.0475  & -5.3086  & 37489923418258428  & True  & 16  & 30  &   0.19  \\
35.0445  & -5.3545  & 35.0444  & -5.3544  & 37490060857210025  & True  & 10  & 91  & -0.05  \\
35.0438  & -5.4894  & 35.0437  & -5.4894  & 37490056562236660  & True  & 75  & 50  &   0.39  \\
35.0436  & -5.4011  & $-$  & $-$  & $-$  & $-$  & 152  & 53  &   0.00  \\
35.0433  & -5.3099  & 35.0432  & -5.3099  & 37490060857213384  & True  & 79  & 67  &   0.13  \\
35.0429  & -5.4777  & 35.0428  & -5.4776  & 37490060857201353  & True  & 58  & 26  &   0.36  \\
\bottomrule
\end{tabular}

 \label{tab:galaxy_properties}
\end{table*}

\bsp	
\label{lastpage}
\end{document}